\documentclass[11pt]{article}
\usepackage{graphicx}
\usepackage{amssymb,amsopn}
\usepackage{amsmath,mathrsfs}
\usepackage{mathtools}
\usepackage{subcaption}
\usepackage[margin=1.25in]{geometry}
\usepackage[usenames,dvipsnames]{color}
\usepackage{url}
\usepackage{flexisym}
\usepackage{comment}
\usepackage[colorlinks = true,
            linkcolor = blue,
            urlcolor  = blue,
            citecolor = blue,
            anchorcolor = blue]{hyperref}
\usepackage{cite}

\newcommand\pubnumber{DESY-22-087}
\newcommand\pubdate{June 2022, revised November 2026}

\def\Title#1{\begin{center} {\LARGE #1 } \end{center}}
\def\Author#1{\begin{center}{ \sc #1} \end{center}}
\def\Address#1{\begin{center}{ \it #1} \end{center}}

\newcommand\pubblock{\rightline{\begin{tabular}{l} \pubnumber\\
         \pubdate \end{tabular}}}
\newenvironment{Abstract}{\begin{quotation} \begin{center}
                       ABSTRACT
     \end{center}\bigskip  }{\end{quotation}}

\def\Acknowledgements{\bigskip  \bigskip \begin{center} \begin{large}
             \bf ACKNOWLEDGEMENTS \end{large}\end{center}}

\begingroup
\catcode`\_11
\catcode`\:11
\gdef\math_bsym_Bin:Nn#1#2{%
\allowbreak\math_char:NNn 2#1{#2}\nobreak}
\gdef\math_bsym_Rel:Nn#1#2{%
\allowbreak\math_char:NNn 3#1{#2}}
\endgroup
\newcommand{\breaktowidth}[2]{\vtop{\hsize#1\noindent$#2$}}
\newcommand{\delimiterswithbreaks}[3]{%
\mathopen{\left#1\vphantom{#3}\right.\hskip-\nulldelimiterspace}
  #3
  \mathclose{\hskip-\nulldelimiterspace\left.\vphantom{#3}\right#2}
	    }

\DeclarePairedDelimiter\abs{\lvert}{\rvert}

\def\Acknowledgements{\bigskip  \bigskip \begin{center} \begin{large}
             \bf ACKNOWLEDGEMENTS \end{large}\end{center}}

\def\beq{\begin{equation}}
\def\eeq#1{\label{#1}\end{equation}}
\def\eeqn{\end{equation}}

\newenvironment{Eqnarray}%
   {\arraycolsep 0.14em\begin{eqnarray}}{\end{eqnarray}}
\def\beqa{\begin{Eqnarray}}
\def\eeqa#1{\label{#1}\end{Eqnarray}}
\def\eeqan{\end{Eqnarray}}

\let\bar=\overbar

\def\bra#1{\left\langle{ #1} \right|}
\def\ket#1{\left| {#1} \right\rangle}

\def\lsim{\mathrel{\raise.3ex\hbox{$<$\kern-.75em\lower1ex\hbox{$\sim$}}}}
\def\gsim{\mathrel{\raise.3ex\hbox{$>$\kern-.75em\lower1ex\hbox{$\sim$}}}}

\def\tr{{ \rm tr}}

\def\tr{{\mbox{\rm tr}}}

\def\del{\partial}
\def\Dslash{\not{\hbox{\kern-4pt $D$}}}
\def\dslash{\not{\hbox{\kern-2pt $\del$}}}
\def\pslash{\not{\hbox{\kern-2pt $p$}}}
\def\ETmiss{\not{\hbox{\kern-4pt $E$}}_T}

\def\Dlr{\mathrel{\raise1.5ex\hbox{$\leftrightarrow$\kern-1em\lower1.5ex\hbox{$D$}}}}

\def\MSB{{\bar{M \kern -2pt S}}}
\def\msb{{\bar{\scriptsize M \kern -1pt S}}}

\def\drb{{\bar{\scriptsize D \kern -1pt R}}}

\def\beq{\begin{equation}}
\def\eeq#1{\label{#1}\end{equation}}
\def\eeqn{\end{equation}}

\def\beqa{\begin{eqnarray}}
\def\eeqa#1{\label{#1}\end{eqnarray}}
\def\eeqan{\end{eqnarray}}

\let\bar=\overbar

\def\bra#1{\left\langle{ #1} \right|}
\def\ket#1{\left| {#1} \right\rangle}

\def\tr{{\mbox{\rm tr}}}

\def\Dslash{\not{\hbox{\kern-4pt $D$}}}
\def\dslash{\not{\hbox{\kern-2pt $\del$}}}

\def\msb{{\bar{\ssstyle M \kern -1pt S}}}

\newcommand{\Nc}{\ensuremath{N_c}}

\newcommand{\SU}{\mathrm{SU}}

\newcommand{\Col}{\ensuremath{\mathbf{c}}}

\newcommand{\ttbar}{t\bar t}
\newcommand{\ttj}{\ifmmode t\bar t+\mathrm{jet} \else $t\bar t+\mathrm{jet}\,\,$\fi}

\newcommand{\Oh}{\mathcal{O}}

\def\msb{\ifmmode \overline{\rm MS}\,\, \else $\overline{\rm MS}\,\, $\fi}

\def\NLO{{\mbox{\scriptsize NLO}}}

\def\virt{{\mbox{\scriptsize virt}}}
\def\real{{\mbox{\scriptsize real}}}
\def\fac{{\mbox{\scriptsize fact}}}
\def\Dipole{{\cal D}}
\def\bra{{\langle}}
\def\ket{{\rangle}}

\def\T#1{\mbox{\bf T}_{#1}}

\begin{document}

\pubblock

\bigskip
\bigskip

\Title{
One-loop soft anomalous dimension matrices for $t\bar t j$ hadroproduction
}

\bigskip

\Author{Bakar Chargeishvili,
  Maria~Vittoria~Garzelli,
  Sven-Olaf~Moch
}


\Address{
  II. Institut f\"ur Theoretische Physik, Universit\"at
  Hamburg, Luruper Chaussee 149, D~--~22761 Hamburg, Germany\\
}

\medskip

\begin{Abstract}
  \noindent
  	We calculate the one-loop corrections to the soft anomalous dimension
	matrices for the production of a top-antitop quark pair in association
	with a jet at hadron colliders. This is a step forward towards implementing
	a procedure for the resummation of soft-gluon emission logarithms for the
	$t\bar{t}j + X$ hadroproduction process, that will enable the improvement
	of the accuracy of the $t\bar{t} j + X$ cross sections, beyond the current
	degree of knowledge. The latter, so far, has reached the next-to-leading
	order (NLO) level, complemented by the accuracy of the Shower Monte Carlo
	approaches used in matching the NLO computations to parton showers (PS) and
	by merging with matrix elements with a different number of light jets.
\end{Abstract}

\vfill
\vfill
\newpage
\setcounter{footnote}{0}

\section{Introduction}
\label{sec:intro}

Studying $t\bar{t}j + X$ production at hadron colliders is one of the relevant
projects for better understanding top-quark physics.
Already some years ago, it was shown that the differential cross sections for
$t\bar{t}j + X$ hadroproduction with respect to the observable $\rho_s=2\,m_t\,
/ \sqrt{s_{\ttbar j}}$ introduced in Ref.~\cite{Alioli:2013mxa},
where $m_t$ is the top-quark mass and $s_{\ttbar j}$ is the squared invariant
mass~of the system of the top, antitop quarks and hardest jet, shows a good
sensitivity to the top-quark mass value and hence can be used for its
determination.
The ATLAS and CMS~collaborations obtained the top-quark mass from the
normalized $\rho_s$ distribution at~$\sqrt{S}=8$~TeV~\cite{Aad:2019mkw,
CMS:2016khu}, following a first analysis by ATLAS at
$\sqrt{S}=7$~TeV~\cite{ATLAS:2014zza, ATLAS:2015pfy}, which served as a
proof-of-concept of the methodology.
Further top-quark mass extractions at $\sqrt{S}=13$~TeV are currently in
preparation and/or in a preliminary status~\cite{CMS:2022qod}, also thanks
to~the most recent high-statistics data on single- and multi-differential
cross sections collected during Run 2 at the Large Hadron Collider (LHC). A
summary of most of the measurements of top-quark mass already performed by the
ATLAS and CMS collaborations, using this and other methods, can be found in
e.g. Ref.~\cite{ATLAS:2022duk}.

Measurements relying on the comparison between Quantum Chromodynamics (QCD) predictions for
(differential) cross sections with a well-defined accuracy and experimental
data, like those mentioned above, are prominent examples of the so-called
indirect methods for determining the top-quark mass. They have become clear
competitors to the direct measurements, based on the kinematical reconstruction
of top quarks from their decay products with the aid of Monte Carlo event
generators, and provide the advantage, with respect to the latter, of leading
to mass values in various well-defined mass renormalization schemes, used for
computing the cross section itself.
Beside the value of the pole mass, the values of the short-distance masses like
the $\overline{\mathrm{MS}}$~\cite{Langenfeld:2009wd},
and the MSR one~\cite{Hoang:2008yj,Hoang:2017suc}, intrinsically more precise
because insensitive to long-distance physics effects, can be
extracted~\cite{Garzelli:2020fmd}.
At present, top-quark mass extractions via the $t\bar{t}j + X$ process are
based on next-to-leading-order (NLO) QCD estimates of
the corresponding production cross sections.
The NLO QCD corrections to $pp \rightarrow t\bar{t}j$ production were first
calculated in Ref.~\cite{Dittmaier:2007wz}.
Uncertainties due to missing higher orders accompanying the NLO QCD
cross sections, translate into uncertainties on the $m_t$ value amounting to
$\sim \pm 0.5$ GeV in the most recent analyses~\cite{CMS:2022qod}. It is
expected that eleva\-ting cross section accuracy beyond NLO could improve the
current indirect estimates of the top-quark mass, leading to reduced systematic
uncertainties. On the other hand, issues in the comparison between
detailed experimental data already available and state-of-the-art theory
predictions with an accuracy not matching the one of the data and, thus,
requiring further improvements, have already been pointed out in case of
multi-differential cross sections for the $t\bar{t}$ production case (see e.g.
Ref.~\cite{CMS:2022uae}).

While NLO QCD computations of $t\bar{t}j$ hadroproduction were matched to
parton shower approaches using various methods~\cite{Kardos:2011qa,
Alioli:2011as, Alwall:2014hca, Czakon:2015cla}, approximate NLO electroweak
corrections were estimated~\cite{Gutschow:2018tuk}, and merging with $t\bar{t}$
computations was also implemented~\cite{Frederix:2012ps, Gutschow:2018tuk}
using fully numerical methods, the extension of the accuracy of both the total
and differential cross sections to the next-to-next-to-leading-order (NNLO) in
QCD still represents significant technical challenges~\cite{Badger:2022mrb}.
At NNLO these consist of the two-loop integrals entering
into double virtual contributions, which are mostly unknown for $2~\rightarrow~3$
kinematics with massive partons, as well as approaches to deal
with the singularities arising from the soft and collinear emissions of
massless particles and with the cancellation of the corresponding infrared
poles, which also need further development.

Meanwhile, in conjunction with the already available amplitudes at LO and NLO,
it is possible to employ the well-established formalism of
resummation~\cite{Catani:1989ne,Sterman:1986aj} of large logarithmic terms
associated to soft-gluon emission, to predict the all-order contribution coming
from this specific class of higher-order terms.
This class provides the dominant contribution to the cross section in the
phase-space regions where the corresponding terms are particularly large, i.e.
close to threshold. By re-expanding the all-order results with a well defined
logarithmic accuracy (leading logarithmic, next-to-leading logarithmic, etc.),
it is then possible to determine differential cross sections with different
degrees of approximate fixed-order accuracy. In particular, approximate NNLO
differential cross sections for the $t\bar{t}j$ hadroproduction process might be obtained for the first time.
One may foresee a scientific development somehow similar to the one
for the $t\bar{t}~+~X$ hadroproduction process,
where, historically, approximate NNLO predictions
for total cross sections~\cite{Moch:2012mk} were published before exact NNLO results could
be computed~\cite{Czakon:2012zr, Czakon:2012pz, Czakon:2013goa,
Catani:2019iny}. The main difference with respect to the $t\bar{t} +X$
hadroproduction process is that, while for the latter it was possible to first
develop calculations for total inclusive cross sections and check the validity
of the threshold approximation, predictions for the $t\bar{t}j + X$
hadroproduction process are divergent already at leading order, implying the
necessity to impose fiducial cuts on the light jet and work at the differential
level from the very beginning.

To be able to successfully apply the resummation formalism at the
next-to-leading lo\-ga\-rithmic (NLL) level, one of the missing ingredients are
the one-loop corrections to the soft functions, that enter the formulas used to
refactorize partonic cross sections in the framework of the resummation
formalism of soft-gluon logarithms.  In the following of this manuscript, we
define the soft functions for  \ttj production in parton-parton scattering and
express them in terms of their building blocks, the soft anomalous dimension
matrices in Section~\ref{sec:4soft}. We provide the details of the calculation
of the latter at 1-loop and the procedure to automatize this step in
Section~\ref{sec:4softcalc}. We report the analytic results of this calculation
in Section~\ref{sec:softdim}. We discuss their connection with the infrared
pole structure of the virtual amplitudes and related cross-checks  in
Section~\ref{sec:poles}. We draw our conclusions in Section~\ref{sec:conclu}.
Technical details on the color bases and the derivation of the color structures
of the soft anomalous dimension matrices are left to the Appendices.
Beyond the specific application on which we focused in this work, a review of
the status of calculations and explicit results on soft anomalous dimension
matrices for a number of processes interesting for collider phenomenology
appeared recently in Ref.~\cite{Kidonakis:2020gxo}. A general recipe for the
calculation of the color structures of these matrices can be found in
Ref.~\cite{Sjodahl:2009wx}. On the other hand, recent developments on the
calculation of soft anomalous dimension matrices in multi-particle scattering
amplitudes at higher orders are summarized in Ref.~\cite{Agarwal:2021ais},
together with up-to-date insights and perspectives for the future.

\section{Definition of the soft function for $t\bar{t}j$ hadroproduction}
\label{sec:4soft}

Factorization theorems allow to express partonic cross sections (or partonic
multi-particle scattering amplitudes) in terms of hard functions, soft
functions and jet functions. Emissions that are neither soft, nor collinear,
are included in the hard functions, collinear non-soft emissions are encoded in
the jet functions and soft wide-angle (i.e. non-collinear) emissions are
described by the soft functions.
The latter kind of emissions are conveniently parametrized by Wilson lines.
Adopting the notation of Ref.~\cite{Kidonakis:1998bk},
a general Wilson line is defined along the  path $\mathcal{C}$ in space-time,
beginning at the point $z^\mu$ and ending at the point $z^{\prime\mu}$:
\begin{equation}
  W^{(f)}\left[\mathcal{C} ; z^{\prime}, z\right]=\mathcal{P} \exp \left[-i g \int_{\eta_{1}}^{\eta_{2}} d \eta \,\, \frac{d y(\eta)}{d \eta} \cdot A^{(f)}(y(\eta))\right] \, ,
\end{equation}
where $g$ is the $\SU(\Nc)$ gauge coupling, with $N_c$ equal to the number of
colors, $A_\mu^{(f)}(y)$ is the gauge field in the fundamental or adjoint
representation of the $\SU(\Nc)$ group, corresponding~to $f$ being a quark or a
gluon, respectively, and $\mathcal{P}$ denotes path-ordering in the same sense
of $\eta$, i.e. when the exponential is expanded, the fields with the higher
value of $\eta$ are on the~left.
The path $\mathcal{C}$ is parametrized by a function $y^\mu(\eta)$ in
the variable $\eta$,
with endpoints $y^\mu(\eta_1)=z^\mu$ and $y^\mu(\eta_2)=z^{\prime\mu}$.
The classical trajectory of an initial or final state parton moving with
velocity $\beta^\mu$ can be expressed as a straight line $y^\mu(\eta)=\eta
\beta^\mu +x^\mu$,
stretching~from the interaction vertex $x^\mu$
to infinity,
either from the distant past
(initial state, $\eta_1 = -\infty$) or towards the distant future
(final state, $\eta_2~=~+\infty$).
The corresponding Wilson line is given by
\begin{equation}
	\Phi_{\beta}^{(f)}\left(\eta_{2}, \eta_{1} ; x\right)=\mathcal{P} \exp \left[-i g \int_{\eta_{1}}^{\eta_{2}} d \eta \,\, \beta \cdot A^{(f)}( \eta \beta + x)\right] \, .
\end{equation}

The $t\bar{t}j$ hadroproduction process has the following structure at the Born
level:
\begin{equation}
  a(\beta_a, c_a^{(f_a)})\,\,b(\beta_b, c_b^{(f_b)}) \to 1(\beta_1, c_1^{(f_1)})\,\,2(\beta_2, c_2^{(f_2)})\,\,3(\beta_3, c_3^{(f_3)}) \, ,
  \label{eq:process}
\end{equation}
where $c^{(f)}_i$ are color indices in the representation $f$ described above,
depending on the kind of parton $i$ under consideration.
In particular, $t\bar{t}j$ hadroproduction at the LHC can occur through
distinct tree-level hard-scattering subprocesses (see
Eq.~(\ref{eq:subprocesses}) in the following).
For each of the two subprocesses $gg \rightarrow t\bar{t}g$ and $q\bar{q} \rightarrow t\bar{t} g$, it is useful to choose a color basis \{$\Col_I$\},
with the index $I$ labeling a generic element of the color basis,
and to define a set of eikonal non-local operators \{$\omega_I$\}, in one by
one correspondence with the elements of the color basis.
We emphasize that the color basis for a specific subprocess is not unique and
depends on the kinds of partons (quarks, antiquarks and gluons) involved in the
subprocess.
In Mellin space we write the $\omega_I$ operator as
\begin{eqnarray}
	\omega_{I}^{\{f\}}(x)_{\left\{c_{k}\right\}}=\sum_{d_{i}} &\Phi_{\beta_{3}}^{f_{3}}(\infty, 0 ; x)_{c_{3}, d_{3}} \Phi_{\beta_{2}}^{f_{2}}(\infty, 0 ; x)_{c_{2}, d_{2}} \Phi_{\beta_{1}}^{f_{1}}(\infty, 0 ; x)_{c_{1}, d_{1}}\left({\Col}_{I}^{\{f\}}\right)_{d_{3} d_{2} d_{1}, d_{b} d_{a}} \nonumber\\
															& \times \Phi_{\beta_{a}}^{f_{a}}(0,-\infty ; x)_{d_{a}, c_{a}} \Phi_{\beta_{b}}^{f_{b}}(0,-\infty ; x)_{d_{b}, c_{b}}\, ,
\end{eqnarray}
which generalizes the expression used in Ref.~\cite{Kidonakis:1998bk} for the
dijet case to our process. In the previous formula the color tensor
$\left({\Col}_{I}^{\{f\}}\right)_{d_{3} d_{2} d_{1}, d_{b} d_{a}}$
links the five Wilson lines corresponding to the external legs of the
subprocess and encodes the coupling of the Wilson lines one with each other in
color space, while the superscript $\{f\}$ refers to the ensemble of the
flavors of all external lines, i.e.  \{$f_a, f_b, f_1, f_2, f_3$\}.
In terms of this operator the eikonal cross section is given
by~\cite{Sterman78,Kidonakis:1996hb}:
\begin{equation}
	\sigma_{L I}^{\{f\}, \mathrm{eik}}\left(\alpha_{s}, \epsilon\right)=\sum_{\xi} \delta(w-w(\xi)) \times\left\langle 0\left|\overline{T}\left[\left(\omega_{L}^{\{f\}}(0)\right)_{\left\{b_{i}\right\}}^{\dagger}\right]\right| \xi\right\rangle\left\langle\xi\left|T\left[\omega_{I}^{\{f\}}(0)_{\left\{b_{i}\right\}}\right]\right| 0\right\rangle,
	\label{eq:4_eik_xsec}
\end{equation}
where $T$ and $\bar{T}$ are time ordering and anti-time ordering operators
respectively, $|\xi\rangle$ denotes a set of intermediate states, $L$ is an
index running over the elements of the color basis, analogously to $I$, and
$\alpha_s = g^2/(4\pi)$ is the strong coupling constant.
The contribution of the $|\xi\rangle$-state to the weight is given by
$w(\xi)$, where $w$ is the corresponding measure of the eikonal phase space
near partonic threshold in the center--of--mass frame of the colliding
partons~\cite{Kidonakis:1998bk}.
In the eikonal approximation of Eq.~(\ref{eq:4_eik_xsec}) soft gluons are
allowed to be collinear as well.
Dimensional regularization is understood,
according to which we work in $D$ dimensions, with $D = 4-\epsilon$, and
soft and soft-collinear singularities are re-expressed
in terms of poles for $\epsilon \rightarrow 0$.
In the definition of the soft function adopted here,
we have to factor out the soft-collinear contributions, since only soft
wide-angle emissions are included\footnote{Soft-collinear emissions
can enter in both the soft and the jet functions, so that different options exist,
all aimed at avoiding double counting.
In the factorization approach applied in this work,
the jet functions include both, collinear non-soft and soft-collinear
emissions, whereas each soft function describes only soft wide-angle gluon emissions (and absorptions)
from the initial and final-state partons~\cite{Collins:1985ue, Sterman:1986aj,
Kidonakis:1998bk}.}.
Thus, the soft function is defined as the part free of collinear divergences of
the eikonal cross section, introduced in Eq.~(\ref{eq:4_eik_xsec}), after a
Mellin transform of the latter,  which leads to the following factorized form
\begin{equation}
  \sigma_{L I}^{\{f\}, e i \mathrm{k}^{N}}=S_{L I}^{\{f\}\,\, N} j_{a}^{N}
  j_{b}^{N} j_{1}^{N} j_{2}^{N} j_{3}^{N} \, .
\end{equation}
In this formula the superscript $N$ indicates that the $N$-th Mellin moment of
the cross section is obtained through the transform,
$S^N$ denotes the soft-function matrix in Mellin space, whereas
$j_{a}^N$, $j_{b}^N$, $j_{1}^N$, $j_{2}^N$, $j_{3}^N$ are the
jet functions corresponding to the initial- and final-state particles,
describing both collinear soft and collinear non-soft emissions.

Notice that, due to this definition, if we expand in a power series the
exponentials which come from the Wilson lines involved in the soft-function
construction, at the Born level, where there is no need for renormalization, we
simply get
\begin{equation}
		S^{\{f\}, 0}_{LI} = \left\langle \Col_L^{\{f\}} | \Col_I^{\{f\}} \right\rangle \, .
		\label{eq:S_born}
\end{equation}

To be able to perform the resummation of the logarithms associated to
soft wide-angle gluon emissions and see how the soft function exponentiates,
we have to study its re\-nor\-ma\-li\-za\-tion properties.
By definition, each soft function is free of collinear divergences.
However, it still contains the ultraviolet (UV) divergences, which can be re-expressed
in terms of poles for $\epsilon \rightarrow 0$ through
dimensional regularization.
Then the UV poles have to be renormalized using appropriate counter-terms.
In the \msb scheme these counter-terms contain the UV divergent
part of the corresponding eikonal amplitudes.
Since the soft function in Eq.~(\ref{eq:4_eik_xsec}) is defined as a product of
two operators, it has to be renormalized multiplicatively:
\begin{equation}
	S_{L I}^{\{f\}^{(B)}}=\left(Z_{S}^{\{f\}^{\dagger}}\right)_{L B} S_{B A}^{\{f\}}\left(Z_{S}^{\{f\}}\right)_{A I} \, ,
	\label{eq:4_soft_reno}
\end{equation}
where the superscript $(B)$ on the left-hand side denotes the bare soft
function and the $Z_{S}$ matrix includes the renormalization counter-terms.

Applying the $\mu\, {d}/{d \mu} $ operation on
Eq.~(\ref{eq:4_soft_reno}) and considering that
 bare quantities do not depend on the renormalization scale $\mu$,
one can derive a renormalization group equation,
which assumes the form~\cite{Kidonakis:1998nf,Botts:1989kf}
\begin{equation}
	\mu \frac{d}{d \mu} S_{L I}^{\{f\}}=\left(\mu \frac{\partial}{\partial \mu}+\beta(\alpha_s) \frac{\partial}{\partial \alpha_s}\right) S_{L I}^{\{f\}}=-\left(\Gamma_{S}^{\{f\}}\right)_{L B}^{\dagger} S_{B I}^{\{f\}}-S_{L A}^{\{f\}}\left(\Gamma_{S}^{\{f\}}\right)_{A I},
	\label{eq:4_soft_rge}
\end{equation}
where $\beta(\alpha_s)$ is the QCD $\beta$-function.
The matrices $\Gamma_S^{\{f\}}$ are process-specific functions of
$g$ through $\alpha_s$. They depend on the process kinematics and color
structure. They are referred to
as \emph{soft anomalous dimension matrices} in the literature.
The solution of the equation above gives the resummation of all logarithms
of the 
scale $\mu$.
By comparing Eq.~(\ref{eq:4_soft_reno}) and Eq.~(\ref{eq:4_soft_rge})
one arrives at the following expression for the soft anomalous dimension
matrices at one-loop in the $\overline{\rm{MS}}$ scheme:
\begin{equation}
  \left(\Gamma_{S}^{\{f\}}\right)_{L I}(\alpha_s) =
  - \alpha_s\, \frac{\partial}{\partial \alpha_s} \operatorname{Res}_{\epsilon \rightarrow 0}\left(Z_{S}^{\{f\}}\right)_{L I}(\alpha_s, \epsilon) \, ,
	\label{eq:4_sad_resid}
\end{equation}
which indicates that soft anomalous dimensions are partial derivatives of
residues of the UV poles contained in the $Z_{S}^{\{f\}}$ matrices.
The solution of Eq.~(\ref{eq:4_soft_rge}) can be written as
\begin{equation}
	S(\mu) = \bar{\mathcal{P}}\exp\left[\int_{\mu_0}^{\mu}\frac{d\mu'}{\mu'}\Gamma_S^\dag(\alpha_s({\mu'}^2))\right]S(\mu_0) {\mathcal{P}} \exp\left[\int_{\mu_0}^{\mu}\frac{d\mu'}{\mu'}\Gamma_S(\alpha_s({\mu'}^2))\right] \, ,
  \label{eq:solution}
\end{equation}
where $\mathcal{P}$ and $\bar{\mathcal{P}}$ refer to path-ordered exponentials
in the same and in the opposite sense as the integration variable $\mu$,
respectively,  and we have suppressed the superscripts $\{f\}$ on top of
$\Gamma_S$ and $S$ to lighten the notation.

Hence, having the soft anomalous dimension matrix, enables us to determine the
soft matrix, provided that this matrix 
is known at an initial scale $\mu_0$. $S(\mu_0)$ can be fixed by matching the
leading-order expansion of the resummed cross section
to the result of a leading-order exact calculation.

The soft anomalous dimension matrices have a perturbative expansion, driven by
the definition of the eikonal cross section, Eq.~(\ref{eq:4_eik_xsec}):
\begin{equation}
	\Gamma_S = \left( \frac{\alpha_s}{2\pi} \right)\Gamma_S^{(1)}+\left( \frac{\alpha_s}{2\pi} \right)^2\Gamma_S^{(2)} + \Oh(\alpha_s^3) .
  \label{eq:pertgamma}
\end{equation}

The result of this expansion can be expressed diagrammatically by means of
\emph{Wilson webs} (for a review on webs see e.g. Ref.~\cite{White:2015wha}),
which are generated in the following way:
\vspace{-0.3cm}
\begin{itemize}
	\item all initial and final state particles are connected to the common
		vertex;

	\item  the external partons, which are involved in the QCD interaction, are
		represented by Wilson lines;

	\item at one-loop level, one has to consider all possible single gluon exchanges
		between the Wilson legs.
\end{itemize}
\vspace{-0.3cm}
The Feynman rules involving the Wilson lines are somewhat different from the
standard QCD Feynman rules involving quarks and gluons and are called
\emph{eikonal Feynman rules}.
In the following we adopt for them the prescriptions of
Ref.~\cite{Kidonakis:1997gm}.

\begin{figure}
	\centering
	\begin{subfigure}[t]{\columnwidth}
		\centering
		\includegraphics[width=0.27\columnwidth]{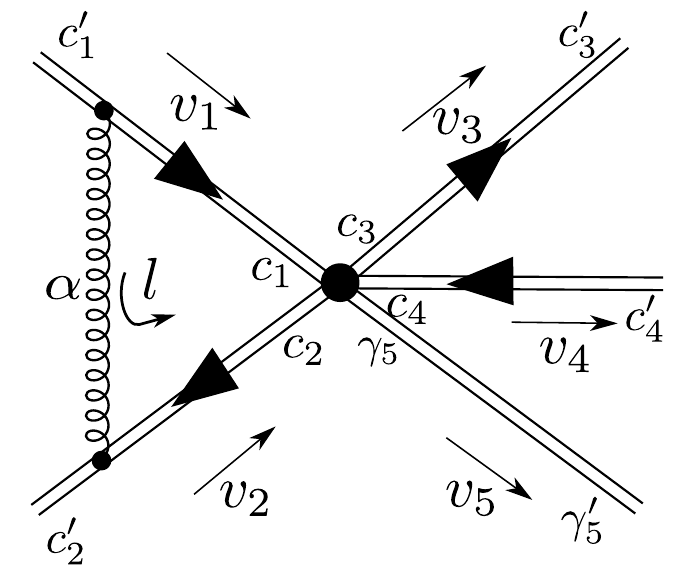}
		\includegraphics[width=0.27\columnwidth]{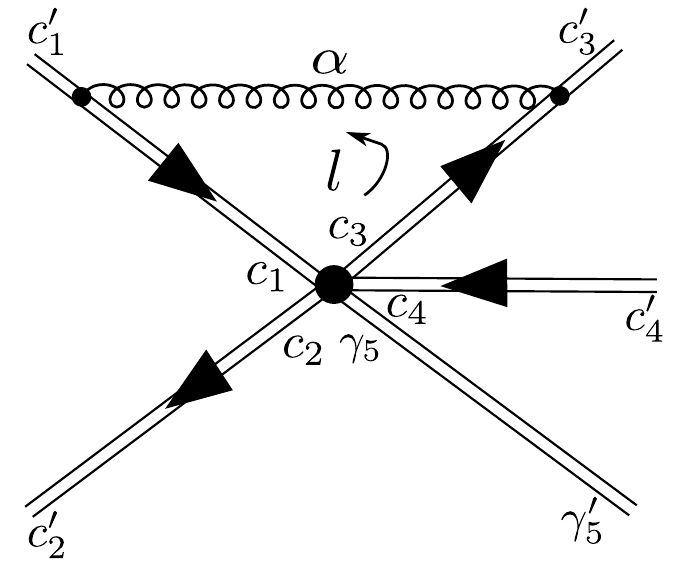}
		\includegraphics[width=0.27\columnwidth]{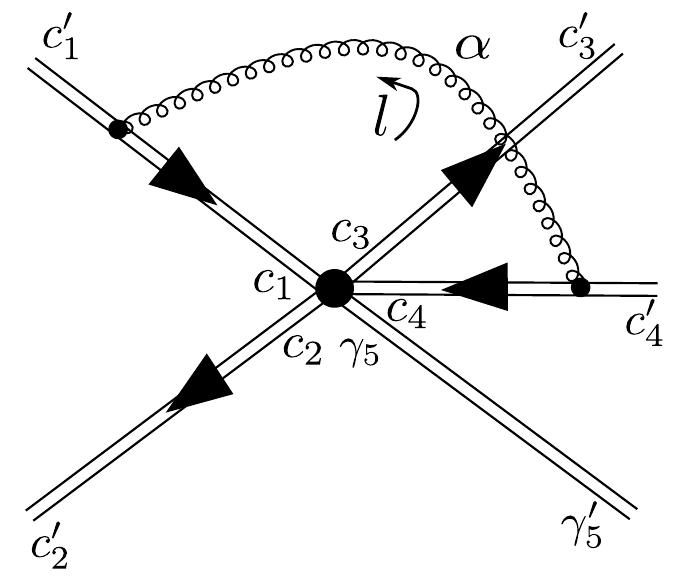}
		\includegraphics[width=0.27\columnwidth]{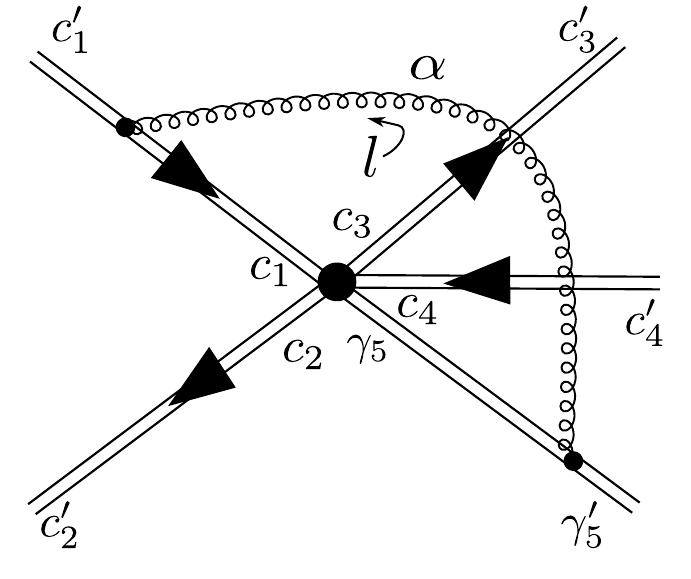}
		\includegraphics[width=0.27\columnwidth]{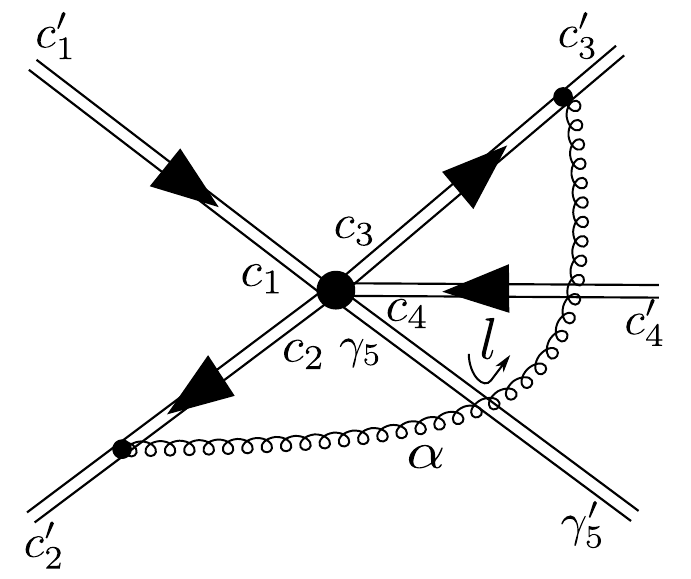}
		\includegraphics[width=0.27\columnwidth]{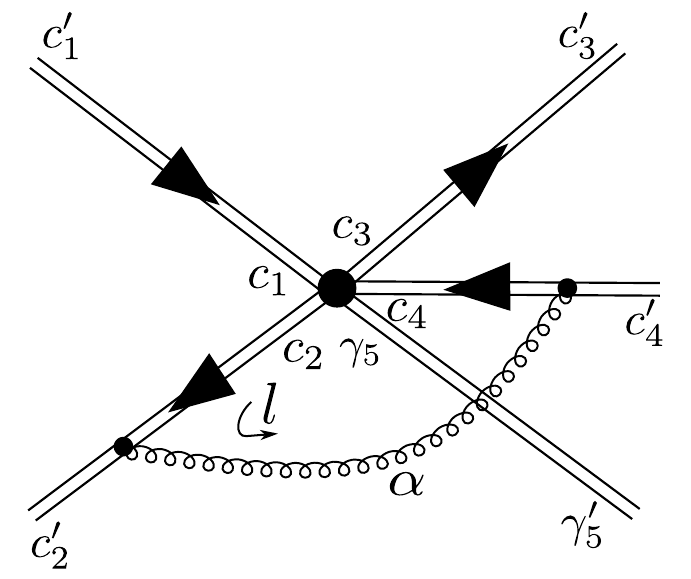}
		\includegraphics[width=0.27\columnwidth]{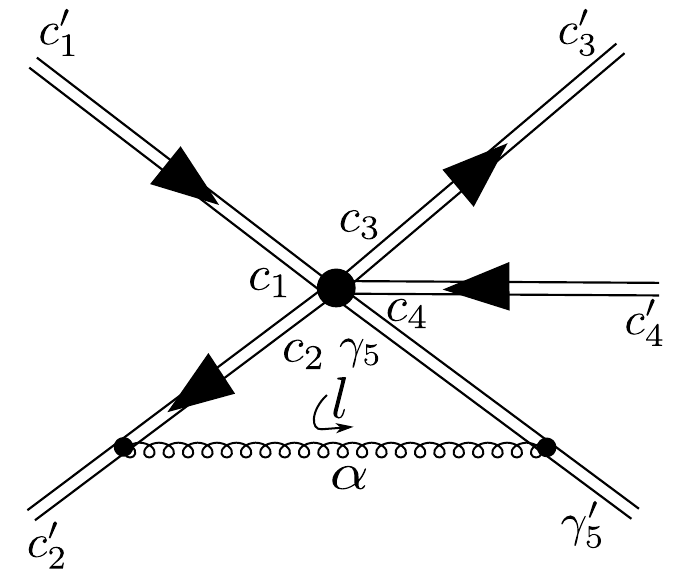}
		\includegraphics[width=0.27\columnwidth]{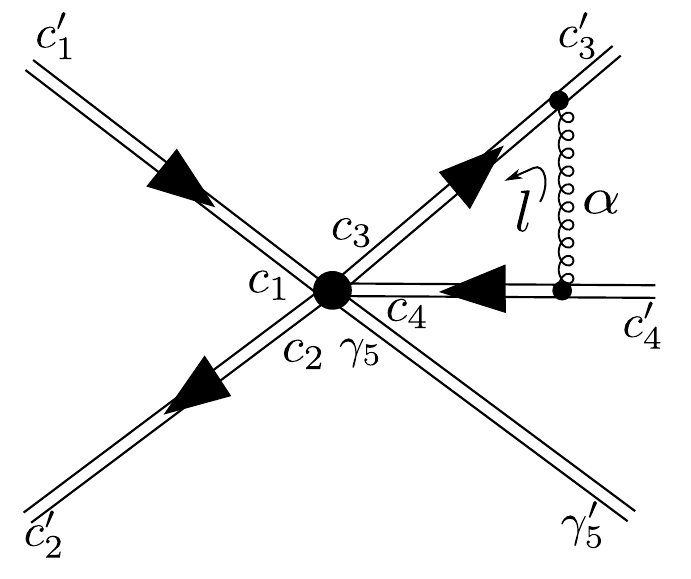}
		\includegraphics[width=0.27\columnwidth]{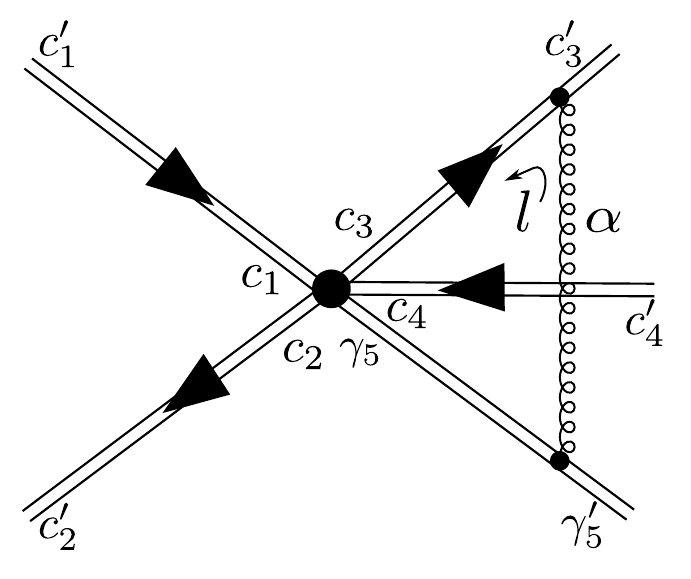}
		\includegraphics[width=0.27\columnwidth]{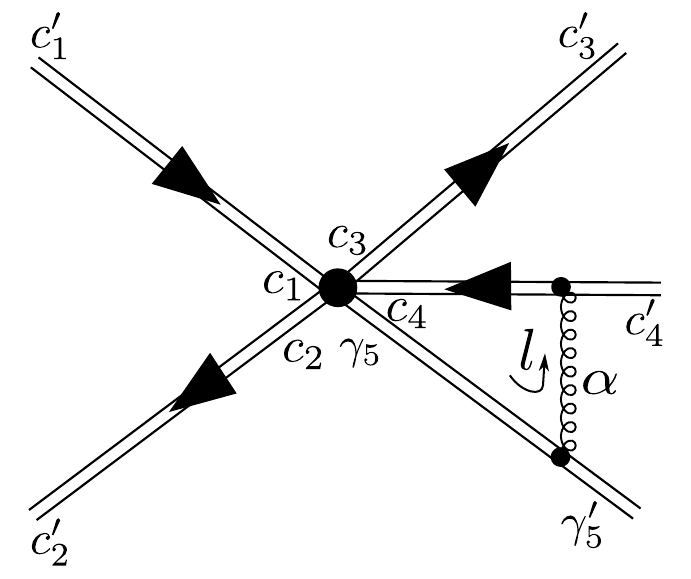}
		\caption{}
		\label{fig:vx_corr}
	\end{subfigure}
	\begin{subfigure}[t]{\columnwidth}
		\centering
		\includegraphics[width=0.27\columnwidth]{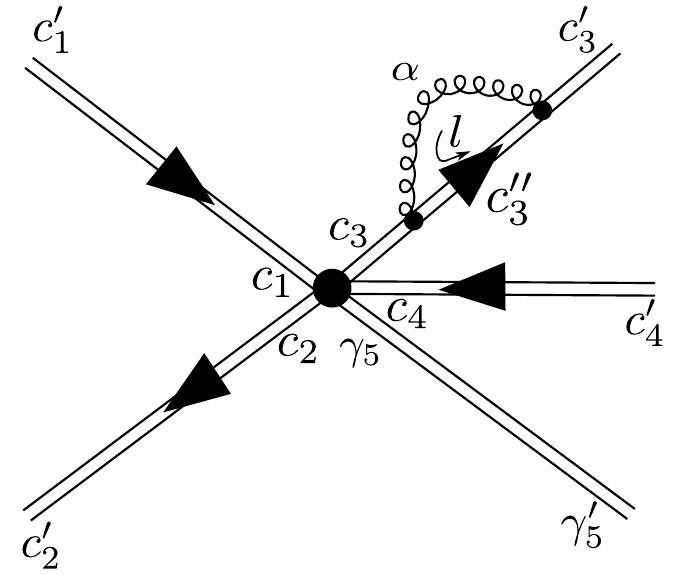}
		\hspace{5em}
		\includegraphics[width=0.27\columnwidth]{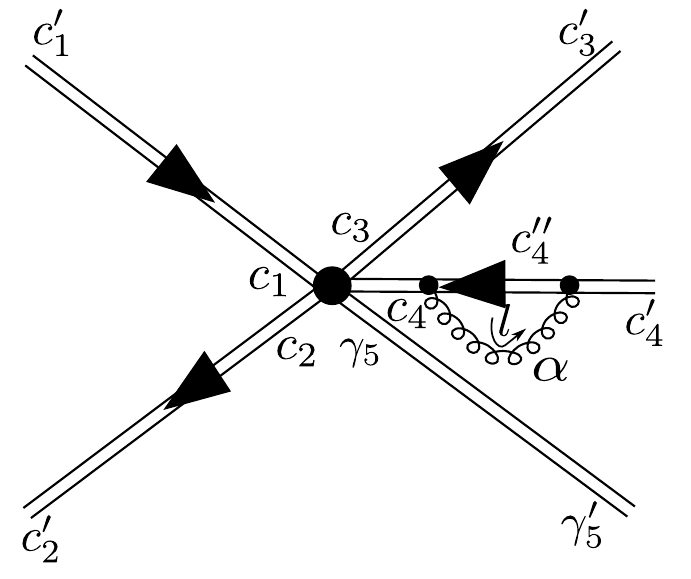}
		\caption{}
		\label{fig:sf_energy}
	\end{subfigure}
	\caption{One-loop corrections to the soft function of the $q\bar{q}
		\rightarrow t\bar{t}g$ partonic subprocess: (a) vertex corrections,
	(b) heavy-quark self-energies. Double lines with arrow correspond to quark
		  Wilson lines, whereas lines wi\-thout arrow correspond to gluon
		  Wilson lines. The direction of a curved arrow shows the momentum flow
		  inside the loop.
		  The Wilson webs for the $g\bar{g} \rightarrow t\bar{t}g$ process can
		  be obtained  by substituting the initial-state quark and antiquark
		  Wilson lines with gluon Wilson lines and leaving all the rest the
		  same.
	}
	\label{fig:graphs}
\end{figure}

\section{Soft anomalous dimension matrices for $t\bar{t}j$ hadroproduction}
\label{sec:4softcalc}

At leading order the hadronic process $p+p \rightarrow \ttj$ consists
of the following partonic subprocesses:
\begin{eqnarray}
&  1)  & q + \bar{q} \rightarrow t + \bar{t} + g \, , \nonumber\\
& 2) & q + g \rightarrow t + \bar{t} + q, \hspace{2em}
            g + q \rightarrow t + \bar{t} + q\, ,  \nonumber\\
&  3) & \bar{q} + g \rightarrow t + \bar{t} + \bar{q}, \hspace{2em}
            g +\bar{q} \rightarrow t + \bar{t} + \bar{q} \, , \nonumber\\
& 4) &  g + g \rightarrow t + \bar{t} + g \, .
            \label{eq:subprocesses}
\end{eqnarray}
Notice that the $t\bar{t}q$ and $t\bar{t}\bar{q}$ production processes is not
symmetric with respect to the exchange of the initial state particles and
hence, we have to consider both the $qg$ and the $gq$ channels explicitly.

From the point of view of color space, processes 2) and 3) do not differ
from process 1).
This is because all these three processes involve the same color vertices.
Hence, we only need to work out processes 1) and 4) explicitly. Results for
processes 2) and 3) can be inferred from those of process 1) by relabeling the
indices.
The color structure of this and other $2 \to 3$ processes has already been
analyzed in Ref.~\cite{Sjodahl:2008fz}.
For our calculations we have used the same color bases presented there and
reported in Appendix~\ref{apx:bases} for completeness.

Let us discuss the case of the $q\bar q\to t\bar t g$ subprocess explicitly.
The same principles apply to the $gg$ induced subprocess.
Figure~\ref{fig:graphs} shows Wilson-web graphs corresponding to all possible
one-loop soft gluon exchanges between the involved partons, plus the
self-energy corrections for the heavy quark and antiquark (on the other hand,
for each massless parton the self-energy contribution vanishes because of the
scaleless loop integral).
The corresponding amplitudes for these graphs, obtained by applying the eikonal
Feynman rules in Ref.~\cite{Kidonakis:1997gm}, are summarized in
Table~\ref{tab:amp}, where we factorize the kinematical ($\kappa_{ij}$) and
color parts ($\mathscr{F}_{ij}$).~Simi\-larly, Table~\ref{tab:amp_gg} reports
the amplitudes for the Wilson web graphs relevant for the
$gg$~$\rightarrow$~$t\bar{t}g$ subprocess. These graphs are analogous to those
reported in Fig.~\ref{fig:graphs}, except for the initial-state Wilson lines
(see caption of Fig.~\ref{fig:graphs}).

\begin{table}[h]
	\centering
\begin{tabular}{c|c|c}
	Connection ($i-j$) & Kinematical part ($\kappa_{ij}$) & Color part ($\mathscr{F}_{ij}$)\\
        $\,$ & before integration & $\,$\\
	\hline
	$1-2$ & $ \frac{v_1^\mu}{-v_1\cdot l+i\epsilon} \frac{-v_2^\nu}{v_2\cdot l+i\epsilon} \frac{N_{\mu\nu}(l)}{l^2+i\epsilon}  $ & $T^\alpha_{c_1c_1'}T^\alpha_{c_2'c_2}\delta_{c_3c_3'}\delta_{c_4c_4'}\delta_{\gamma_5\gamma_5'}$ \\
	$1-3$ & $ \frac{v_1^\mu}{v_1\cdot l+i\epsilon} \frac{v_3^\nu}{v_3\cdot l+i\epsilon} \frac{N_{\mu\nu}(l)}{l^2+i\epsilon}  $ & $T^\alpha_{c_1c_1'}T^\alpha_{c_3'c_3}\delta_{c_2c_2'}\delta_{c_4c_4'}\delta_{\gamma_5\gamma_5'}$ \\
	$1-4$ & $ \frac{v_1^\mu}{v_1\cdot l+i\epsilon} \frac{-v_4^\nu}{v_4\cdot l+i\epsilon} \frac{N_{\mu\nu}(l)}{l^2+i\epsilon}  $  & $T^\alpha_{c_1c_1'}T^\alpha_{c_4c_4'}\delta_{c_2c_2'}\delta_{c_3c_3'}\delta_{\gamma_5\gamma_5'}$ \\
	$1-5$ & $ \frac{v_1^\mu}{v_1\cdot l+i\epsilon} \frac{-v_5^\nu}{v_5\cdot l+i\epsilon} \frac{N_{\mu\nu}(l)}{l^2+i\epsilon}  $  & $T^\alpha_{c_1c_1'}(-if^{\alpha\gamma_5\gamma_5'})\delta_{c_2c_2'}\delta_{c_3c_3'}\delta_{c_4c_4'}$ \\
	$2-3$ & $ \frac{-v_2^\mu}{-v_2\cdot l+i\epsilon} \frac{v_3^\nu}{-v_3\cdot l+i\epsilon} \frac{N_{\mu\nu}(l)}{l^2+i\epsilon} $ & $T^\alpha_{c_2'c_2}T^\alpha_{c_3'c_3}\delta_{c_1c_1'}\delta_{c_4c_4'}\delta_{\gamma_5\gamma_5'}$ \\
	$2-4$ & $ \frac{-v_2^\mu}{-v_2\cdot l+i\epsilon} \frac{-v_4^\nu}{-v_4\cdot l+i\epsilon} \frac{N_{\mu\nu}(l)}{l^2+i\epsilon}$ & $T^\alpha_{c_2'c_2}T^\alpha_{c_4c_4'}\delta_{c_1c_1'}\delta_{c_3c_3'}\delta_{\gamma_5\gamma_5'}$ \\
	$2-5$ & $ \frac{-v_2^\mu}{-v_2\cdot l+i\epsilon} \frac{v_5^\nu}{-v_5\cdot l+i\epsilon} \frac{N_{\mu\nu}(l)}{l^2+i\epsilon}$ & $T^\alpha_{c_2'c_2}(-if^{\alpha\gamma_5'\gamma_5})\delta_{c_1c_1'}\delta_{c_3c_3'}\delta_{c_4'c_4}$ \\
	$3-4$ & $ \frac{v_3^\mu}{-v_3\cdot l+i\epsilon} \frac{-v_4^\nu}{v_4\cdot l+i\epsilon} \frac{N_{\mu\nu}(l)}{l^2+i\epsilon}  $ & $T^\alpha_{c_3'c_3}T^\alpha_{c_4c_4'}\delta_{c_1c_1'}\delta_{c_2c_2'}\delta_{\gamma_5\gamma_5'}$ \\
	$3-5$ & $ \frac{v_3^\mu}{-v_3\cdot l+i\epsilon} \frac{-v_5^\nu}{v_5\cdot l+i\epsilon} \frac{N_{\mu\nu}(l)}{l^2+i\epsilon}  $ & $T^\alpha_{c_3'c_3}(-if^{\alpha\gamma_5\gamma_5'})\delta_{c_1c_1'}\delta_{c_2c_2'}\delta_{c_4c_4'}$ \\
	$4-5$ & $ \frac{v_4^\mu}{-v_4\cdot l+i\epsilon} \frac{-v_5^\nu}{v_5\cdot l+i\epsilon} \frac{N_{\mu\nu}(l)}{l^2+i\epsilon} $ & $T^\alpha_{c_4c_4'}(-if^{\alpha\gamma_5\gamma_5'})\delta_{\gamma_1\gamma_1'}\delta_{\gamma_2\gamma_2'}\delta_{c_3c_3'}$ \\
	$3-3$ & $ \frac{v_3^\mu}{v_3\cdot l+i\epsilon} \frac{v_3^\nu}{-v_3\cdot l+i\epsilon} \frac{N_{\mu\nu}(l)}{l^2+i\epsilon} $ & $T^\alpha_{c_3''c_3}T^\alpha_{c_3'c_3''}\delta_{c_1c_1'}\delta_{c_2c_2'}\delta_{c_4c_4'}\delta_{\gamma_5\gamma_5'}$ \\
	$4-4$ & $ \frac{-v_4^\mu}{v_4\cdot l+i\epsilon} \frac{-v_4^\nu}{-v_4\cdot l+i\epsilon} \frac{N_{\mu\nu}(l)}{l^2+i\epsilon} $ & $T^\alpha_{c_4''c_4}T^\alpha_{c_4'c_4''}\delta_{c_1c_1'}\delta_{c_2c_2'}\delta_{c_3c_3'}\delta_{\gamma_5\gamma_5'}$ \\
\end{tabular}
\caption{Amplitudes for the graphs shown in Fig.~\ref{fig:graphs}. The first
	column specifies the partons which exchange a soft gluon, the second and
	third column show the corresponding kinematic part $\kappa_{ij}$ and color
	part $\mathscr{F}_{ij}$.  For the definition of the various elements
	appearing in the color part see Appendix~\ref{apx:bases}.  The first ten
	rows correspond to vertex corrections (Fig.~\ref{fig:vx_corr}), whereas the
	last two rows correspond to the self-energy contributions
	(Fig.~\ref{fig:sf_energy}).
}
\label{tab:amp}
\end{table}

\begin{table}[h]
	\centering
\begin{tabular}{c|c|c}
  Connection ($i-j$) & Kinematical part ($\kappa_{ij}$) & Color part ($\mathscr{F}_{ij}$)\\
         $\,$ & before integration & $\,$\\
	\hline
	$1-2$ & $ \frac{v_1^\mu}{-v_1\cdot l+i\epsilon} \frac{-v_2^\nu}{v_2\cdot l+i\epsilon} \frac{N_{\mu\nu}(l)}{l^2+i\epsilon}  $ & $-f^{\alpha \gamma_1\gamma_1'}f^{\alpha \gamma_2'\gamma_2}\delta_{c_3c_3'}\delta_{c_4c_4'}\delta_{\gamma_5\gamma_5'}$ \\
	$1-3$ & $ \frac{-v_1^\mu}{v_1\cdot l+i\epsilon} \frac{v_3^\nu}{v_3\cdot l+i\epsilon} \frac{N_{\mu\nu}(l)}{l^2+i\epsilon}  $ & $(-if^{\alpha \gamma_1'\gamma_1})T^\alpha_{c_3'c_3}\delta_{\gamma_2\gamma_2'}\delta_{c_4c_4'}\delta_{\gamma_5\gamma_5'}$ \\
	$1-4$ & $ \frac{-v_1^\mu}{v_1\cdot l+i\epsilon} \frac{-v_4^\nu}{v_4\cdot l+i\epsilon} \frac{N_{\mu\nu}(l)}{l^2+i\epsilon}  $  & $(-if^{\alpha \gamma_1'\gamma_1})T^\alpha_{c_4c_4'}\delta_{\gamma_2\gamma_2'}\delta_{c_3c_3'}\delta_{\gamma_5\gamma_5'}$ \\
	$1-5$ & $ \frac{-v_1^\mu}{v_1\cdot l+i\epsilon} \frac{-v_5^\nu}{v_5\cdot l+i\epsilon} \frac{N_{\mu\nu}(l)}{l^2+i\epsilon}  $  & $-f^{\alpha \gamma_1'\gamma_1}f^{\alpha\gamma_5\gamma_5'}\delta_{\gamma_2\gamma_2'}\delta_{c_3c_3'}\delta_{c_4c_4'}$ \\
	$2-3$ & $ \frac{v_2^\mu}{-v_2\cdot l+i\epsilon} \frac{v_3^\nu}{-v_3\cdot l+i\epsilon} \frac{N_{\mu\nu}(l)}{l^2+i\epsilon} $ & $(-if^{\alpha \gamma_2\gamma_2'})T^\alpha_{c_3'c_3}\delta_{\gamma_1\gamma_1'}\delta_{c_4c_4'}\delta_{\gamma_5\gamma_5'}$ \\
	$2-4$ & $ \frac{v_2^\mu}{-v_2\cdot l+i\epsilon} \frac{-v_4^\nu}{-v_4\cdot l+i\epsilon} \frac{N_{\mu\nu}(l)}{l^2+i\epsilon}$ & $(-if^{\alpha \gamma_2\gamma_2'})T^\alpha_{c_4c_4'}\delta_{\gamma_1\gamma_1'}\delta_{c_3c_3'}\delta_{\gamma_5\gamma_5'}$ \\
	$2-5$ & $ \frac{v_2^\mu}{-v_2\cdot l+i\epsilon} \frac{v_5^\nu}{-v_5\cdot l+i\epsilon} \frac{N_{\mu\nu}(l)}{l^2+i\epsilon}$ & $-f^{\alpha \gamma_2\gamma_2'}f^{\alpha\gamma_5'\gamma_5}\delta_{c_1c_1'}\delta_{c_3c_3'}\delta_{c_4'c_4}$ \\
	$3-4$ & $ \frac{v_3^\mu}{-v_3\cdot l+i\epsilon} \frac{-v_4^\nu}{v_4\cdot l+i\epsilon} \frac{N_{\mu\nu}(l)}{l^2+i\epsilon}  $ & $T^\alpha_{c_3'c_3}T^\alpha_{c_4c_4'}\delta_{\gamma_1\gamma_1'}\delta_{\gamma_2\gamma_2'}\delta_{\gamma_5\gamma_5'}$ \\
	$3-5$ & $ \frac{v_3^\mu}{-v_3\cdot l+i\epsilon} \frac{-v_5^\nu}{v_5\cdot l+i\epsilon} \frac{N_{\mu\nu}(l)}{l^2+i\epsilon}  $ & $T^\alpha_{c_3'c_3}(-if^{\alpha\gamma_5\gamma_5'})\delta_{\gamma_1\gamma_1'}\delta_{\gamma_2\gamma_2'}\delta_{c_4c_4'}$ \\
	$4-5$ & $ \frac{-v_4^\mu}{-v_4\cdot l+i\epsilon} \frac{-v_5^\nu}{v_5\cdot l+i\epsilon} \frac{N_{\mu\nu}(l)}{l^2+i\epsilon} $ & $T^\alpha_{c_4c_4'}(-if^{\alpha\gamma_5\gamma_5'})\delta_{\gamma_1\gamma_1'}\delta_{\gamma_2\gamma_2'}\delta_{c_3c_3'}$ \\
	$3-3$ & $ \frac{v_3^\mu}{v_3\cdot l+i\epsilon} \frac{v_3^\nu}{-v_3\cdot l+i\epsilon} \frac{N_{\mu\nu}(l)}{l^2+i\epsilon} $ & $T^\alpha_{c_3''c_3}T^\alpha_{c_3'c_3''}\delta_{\gamma_1\gamma_1'}\delta_{\gamma_2\gamma_2'}\delta_{c_4c_4'}\delta_{\gamma_5\gamma_5'}$ \\
	$4-4$ & $ \frac{-v_4^\mu}{v_4\cdot l+i\epsilon} \frac{-v_4^\nu}{-v_4\cdot l+i\epsilon} \frac{N_{\mu\nu}(l)}{l^2+i\epsilon} $ & $T^\alpha_{c_4''c_4}T^\alpha_{c_4'c_4''}\delta_{c_1c_1'}\delta_{c_2c_2'}\delta_{c_3c_3'}\delta_{\gamma_5\gamma_5'}$ \\
\end{tabular}
\caption{
	Same as Table~\ref{tab:amp}, but for the $gg\rightarrow t\bar t g$ subprocess.
}
\label{tab:amp_gg}
\end{table}

We work in a general axial gauge, introducing a vector $n^\mu$, whose
components act as gauge parameters, which satisfies the conditions $n \cdot A =
0$, $n^2 \ne 0$. This simplifies the analysis, by ensuring that collinear
logarithms appear only within the jet functions~\cite{Kidonakis:1998bk}.  In
this gauge the gluon propagator entering the kinematical factors
$\kappa_{ij}$ is given by
\begin{equation}
	N^{\mu\nu}(k)=g^{\mu\nu}-\frac{n^\mu k^\nu+n^\nu k^\mu}{n\cdot k}+n^2\frac{k^\mu k^\nu}{(n\cdot k)^2} \, ,
\end{equation}
where $g^{\mu\nu}$ is the metric tensor.
To deal with the unphysical singularities introduced by the axial gauge
we use the principal value prescription~\cite{Leibbrandt}:
\begin{equation}
	\frac{\mathscr{P}}{(l\cdot n)^\beta}=\frac{1}{2}\left(\frac{1}{(l\cdot n+i\epsilon)^\beta}+(-1)^{\beta}\frac{1}{(-l\cdot n + i\epsilon)^\beta}\right)\, ,
\end{equation}
where $\beta$ denotes a power not restricted to integer values.
As a result, each integral over the kinematic part can be reduced to the
following form:
\begin{align}
	\omega_{ij}(\delta_iv_i, \delta_jv_j,\Delta_i,\Delta_j) &= g^2\int\frac{d^{D}\ell}{(2\pi)^D} \kappa_{ij}(\delta_iv_i, \delta_jv_j,\Delta_i,\Delta_j;\ell)\nonumber\\
															&=\Delta_i\Delta_j\delta_i\delta_j\Bigg(I_1(\delta_iv_i,\delta_jv_j)-\frac{1}{2}I_2(\delta_iv_i,n)-\frac{1}{2}I_2(\delta_iv_i,-n) - \nonumber\\
															&\qquad\qquad\qquad-\frac{1}{2}I_3(\delta_jv_j,n)-\frac{1}{2}I_3(\delta_jv_j,-n)+I_4(n^2) \Bigg) \, ,
	\label{eq:int_master}
\end{align}
where $v^\mu_i$ is the dimensionless velocity vector of the parton $i$ with
momentum $p^\mu_i$, defined as:
\begin{equation}
  v_i^\mu = \frac{p_i^\mu}{Q}\, ,
\label{eq:v_i}
\end{equation}
with $Q=\sqrt{s/2}$, where $s$ is the partonic center-of-mass energy.
A term $\Delta_i$ is associated to each quark and antiquark, assuming values
of $+ 1$ and $-1$, respectively. In case of gluons $\Delta_i = 1$ if a soft
gluon line is emitted below the eikonal line and $\Delta_i = -1$
otherwise~\cite{Kidonakis:1998nf}.
The term $\delta_i$  assumes value $+1$ or $-1$
depending if the direction of the momentum of the exchanged gluon is the same
or opposite of the direction of momentum of line $i$.
More details on $\Delta_i$ and $\delta_i$, as well as on the integrals $I_1$,
$I_2$, $I_3$ and $I_4$ in Eq.~(\ref{eq:int_master}),
can be found in Ref.~\cite{Kidonakis:1997gm}.

Considering the one-loop soft-anomalous dimension expression in
Eq.~(\ref{eq:4_sad_resid}), we are only interested in the simple poles of the
integrals above.
The UV poles of the integrals $I_1~-~I_4$ appearing in
Eq.~(\ref{eq:int_master}) are given in Ref.~\cite{Kidonakis:1997gm} for $2\to2$
processes. Two of the expressions for $I_1$ have to be generalized to the present
$2\to3$ kinematics. By Lorentz invariance, $I_1(\delta_iv_i,\delta_jv_j)$ depends on
the velocities only through $v_i^2$, $v_j^2$ and $\delta_i\delta_j\,v_{ij}$, with
$v_{ij}=v_i\cdot v_j$, where $\delta_i\delta_j=-1$ if the partons $i$ and $j$ are
both incoming or both outgoing, and $\delta_i\delta_j=+1$ otherwise.
For two massive partons the pole of $I_1$ is proportional to the function $L_\beta$
given below. It depends on the invariant mass of the pair, which for the top-quark
pair is $s_{t\bar t}=(p_3+p_4)^2$ and coincides with $s$ only for $2\to2$ processes.
For one massive parton of mass $m$ and one massless parton, Ref.~\cite{Kidonakis:1997gm}
considers an incoming massless and an outgoing massive parton, and the logarithm in
the pole of $I_1$ reads $\ln(v_{ij}^2\,s/(2m^2))$. This form conceals the sign of
$\delta_i\delta_j$. In general the logarithm is
$2\ln(\delta_i\delta_j\,v_{ij})+\ln(s/(2m^2))$, which differs from the expression
above for the dipoles formed by the outgoing jet and the outgoing top quark or antiquark.
As in the massless case of Ref.~\cite{Kidonakis:1997gm}, where the argument of the
logarithm is $\delta_i\delta_j\,v_{ij}/2$, the $i\epsilon$ prescription of the
eikonal propagators implies $\ln(-x)=\ln x-i\pi$ for $x>0$.
One can observe that plugging these integrals into Eq.~(\ref{eq:int_master})
causes double poles to cancel.
There are two sources of gauge dependence in these integrals:
i) the gauge dependence coming from the vertex-correction webs, among
those in Fig.~\ref{fig:vx_corr}, where the wide-angle soft gluon is exchanged
between a massive quark and another parton, is canceled by the heavy-quark
self-energy contributions depicted in Fig.~\ref{fig:sf_energy}; ii)
the remaining gauge dependence originates from the massless partons and it
is contained in terms of the following form
\begin{equation}
  \nu_i= \frac{(v_i\cdot n)^2}{\abs{n}^2} \, .
  \end{equation}
They remain present at the level of the soft function and are expected to
finally cancel upon substitution of the soft function into the fully factorized
expression of the partonic cross section,
thanks to their combination with similar terms coming from the jet
functions, as also discussed in Ref.~\cite{Kidonakis:1997gm}.

The final result for the soft anomalous dimension matrix elements
can be written as:
\begin{equation}
  (\Gamma_S)^{AB} =\sum_{i,j=1;i\leq j}^{5} \mathcal{F}^{AB}_{ij}\Omega_{ij} \, ,
  \label{eq:soft_to}
\end{equation}
where the sum runs over all diagrams of Fig.~\ref{fig:graphs},
$\mathcal{F}_{ij}$ are color matrices, corresponding to Wilson webs with soft
gluon exchange between partons $i$ and $j$, $A$ and $B$ are indices running
over the elements of the color basis and $\Omega_{ij} =
(\omega_{ij})_{\mathrm{UV~divergent}}$ corresponds to the sum of simple poles
of $\omega_{ij}$ (see Eq.~(\ref{eq:int_master})), extracted using the technique
described above.

In Eq.~(\ref{eq:soft_to}) the self-energy contributions of
Fig.~\ref{fig:sf_energy} enter with a prefactor $\frac{1}{2}$, due to the
way the quark (antiquark) wave-function renormalization is performed:
\begin{equation}
	\psi^{(B)}=Z_\psi^{1/2}\psi \, .
\end{equation}

Concerning the color decomposition of the amplitude associated to each Wilson
web in the color basis, in simple cases, as the one discussed in
Appendix~\ref{apx:col_dec}, one is able to manipulate the color tensor
structure and visually re\-co\-gnize  linear combinations in terms of the
elements of the color basis, which allow for a manual reconstruction of
$\mathcal{F}_{ij}$.
However, for more complicated cases, like those for the $gg\rightarrow
t\bar{t}g$ subprocess, requiring a color basis with many elements, it becomes
hard to express the color structure in terms of the basis tensors in a
straightforward way.
Hence, we have developed a more general and automatized approach to achieve this
task. Let us define the following tensor:
\begin{equation}
	G_{AB} = \tr\left(\Col_A{\Col_B^\prime}^\dag\sum_{i,j;i\leq j}\Omega_{ij}\mathscr{F}_{ij}\right),
\end{equation}
where $\{\Col \}$ is the color basis for the process before soft wide-angle
gluon exchange, $\{\Col^\prime \}$ is the color basis in terms of color indices
after soft wide-angle gluon exchange, the $\mathscr{F}_{ij}$ terms are taken
directly from the Wilson web graphs and collected in Table~\ref{tab:amp}
and~\ref{tab:amp_gg}, and the trace is performed over color space.
The idea of this procedure is to contract the open color indices and
algebraically manipulate the resulting terms which are scalars in color space.
This way we perform the color decomposition of the amplitude of each Wilson web.
Then, the components of the soft anomalous dimension can be extracted according
to
\begin{equation}
	\Gamma_{IJ} = \left(S^{(0)}\right)^{-1}_{\,IK}\,\, G_{KJ} \, .
  \label{eq:get_gamma}
\end{equation}
These operations can be implemented using computer algebra systems, which makes
it possible to perform the calculations even for complicated processes.


\section{Results for one-loop soft anomalous dimension matrices}
\label{sec:softdim}

Following the strategy outlined in Section~\ref{sec:4softcalc} and using the
computer algebra system {\tt FORM}~\cite{FORM4}, the package {\tt
  FORMSoft}~\cite{formsoft} has been developed to perform
analytical calculations of soft anomalous dimension matrices at one loop.
Starting from color bases hard-coded by the user, the rest is done
automatically.  The generation of Wilson webs and of their amplitudes, which
are input for {\tt FORMSoft}, has been performed automatically as well, using
the {\tt C}-program {\tt WilsonWebs}~\cite{wilsonwebs}, which supports  planar
Wilson web generation up to six loops.
In the following we present explicitly the analytical results of the
calculation for the $q\bar{q}$- and $gg$-induced subprocesses.

\subsection{$q\bar{q}$ channel}
\label{sec:qqchannel}

The Born-level soft matrix calculated according to Eq.~(\ref{eq:S_born}) is
given by
\begin{equation}
	S^{(0)} = \begin{pmatrix}
		\frac{\Nc(\Nc^2-1)}{2} & 0 & 0 & 0\\
		0& \frac{\Nc(\Nc^2-1)}{2} & 0 & 0\\
		0& 0& \frac{\Nc(\Nc^2-1)}{4} & 0\\
		0& 0& 0& \frac{(\Nc^2-1)(\Nc^2-4)}{4\Nc} \\
        \end{pmatrix}
        \, .
\end{equation}
We explicitly verified that $S^{(0)}$ satisfies the consistency check
\begin{equation}
  tr(H^{(0)}_{AB}S^{(0)}_{BC}) = |\mathcal{M}|^2 \, ,
  \label{eq:check0}
\end{equation}
where $H^{(0)}$ is the color decomposed Born-level hard-scattering matrix
entering the refacto\-ri\-zation formula for the partonic cross section and
$|\mathcal{M}|^2$ is the squared amplitude of the partonic process at
leading order.
Performing the calculation of the color part of the soft anomalous dimension
matrix  at one-loop, we are able to reproduce the results published
in~\cite{Sjodahl:2008fz}, confirming their correctness, using their same color
basis, also reported in Appendix~\ref{apx:basis_qq} for completeness.
After including the kinematical factors as well, we get the following
results for the elements of the one-loop $\Gamma$ matrix, considering the
perturbative expansion in Eq.~(\ref{eq:pertgamma}):

{\small{
\setlength{\jot}{10pt}
\allowdisplaybreaks
\begin{align*}
	\Gamma^{(1)}_{1,1} =&\frac{1}{2 N_{c}}\breaktowidth{0.8\linewidth}{\delimiterswithbreaks{[}{]}{2 L_{\beta} + N_{c}^{2} \delimiterswithbreaks{(}{)}{- 2 \log{\delimiterswithbreaks{(}{)}{\nu_{5}}} + 2 \log{\delimiterswithbreaks{(}{)}{- v_{35}}} + 2 \log{\delimiterswithbreaks{(}{)}{- v_{45}}} + 2 \log{\delimiterswithbreaks{(}{)}{\frac{s}{m_{t}^{2}}}} - \log{\delimiterswithbreaks{(}{)}{16}} + 2} + \delimiterswithbreaks{(}{)}{N_{c}^{2} - 1} \delimiterswithbreaks{(}{)}{- \log{\delimiterswithbreaks{(}{)}{\nu_{1}}} - \log{\delimiterswithbreaks{(}{)}{\nu_{2}}} - 2 i \pi} + \log{\delimiterswithbreaks{(}{)}{4}}}}\nonumber \\
\Gamma^{(1)}_{1,2} =&\frac{1}{N_{c}}\breaktowidth{0.8\linewidth}{\delimiterswithbreaks{[}{]}{\log{\delimiterswithbreaks{(}{)}{v_{13}}} - \log{\delimiterswithbreaks{(}{)}{v_{14}}} - \log{\delimiterswithbreaks{(}{)}{v_{23}}} + \log{\delimiterswithbreaks{(}{)}{v_{24}}}}}\nonumber \\
\Gamma^{(1)}_{1,3} =&\breaktowidth{0.8\linewidth}{- \frac{\log{\delimiterswithbreaks{(}{)}{v_{13}}}}{2} - \frac{\log{\delimiterswithbreaks{(}{)}{v_{14}}}}{2} + \log{\delimiterswithbreaks{(}{)}{v_{15}}} + \frac{\log{\delimiterswithbreaks{(}{)}{v_{23}}}}{2} + \frac{\log{\delimiterswithbreaks{(}{)}{v_{24}}}}{2} - \log{\delimiterswithbreaks{(}{)}{v_{25}}}}\nonumber \\
\Gamma^{(1)}_{1,4} =&\frac{N_{c}^{2} - 4}{2 N_{c}^{2}}\breaktowidth{0.8\linewidth}{\delimiterswithbreaks{[}{]}{\log{\delimiterswithbreaks{(}{)}{v_{13}}} - \log{\delimiterswithbreaks{(}{)}{v_{14}}} - \log{\delimiterswithbreaks{(}{)}{v_{23}}} + \log{\delimiterswithbreaks{(}{)}{v_{24}}}}}\nonumber \\
\Gamma^{(1)}_{2,1} =&\frac{1}{N_{c}}\breaktowidth{0.8\linewidth}{\delimiterswithbreaks{[}{]}{\log{\delimiterswithbreaks{(}{)}{v_{13}}} - \log{\delimiterswithbreaks{(}{)}{v_{14}}} - \log{\delimiterswithbreaks{(}{)}{v_{23}}} + \log{\delimiterswithbreaks{(}{)}{v_{24}}}}}\nonumber \\
\Gamma^{(1)}_{2,2} =&\frac{1}{2 N_{c}}\breaktowidth{0.8\linewidth}{\delimiterswithbreaks{[}{]}{N_{c}^{2} \delimiterswithbreaks{(}{)}{- 2 \log{\delimiterswithbreaks{(}{)}{\nu_{5}}} + 2 \log{\delimiterswithbreaks{(}{)}{v_{15}}} + 2 \log{\delimiterswithbreaks{(}{)}{v_{25}}} - \log{\delimiterswithbreaks{(}{)}{16}} + 2} + \delimiterswithbreaks{(}{)}{N_{c}^{2} - 1} \delimiterswithbreaks{(}{)}{- 2 L_{\beta} - \log{\delimiterswithbreaks{(}{)}{\nu_{1}}} - \log{\delimiterswithbreaks{(}{)}{\nu_{2}}}} + \log{\delimiterswithbreaks{(}{)}{4}} + 2 i \pi}}\nonumber \\
\Gamma^{(1)}_{2,3} =&\breaktowidth{0.8\linewidth}{- \frac{\log{\delimiterswithbreaks{(}{)}{v_{13}}}}{2} + \frac{\log{\delimiterswithbreaks{(}{)}{v_{14}}}}{2} - \frac{\log{\delimiterswithbreaks{(}{)}{v_{23}}}}{2} + \frac{\log{\delimiterswithbreaks{(}{)}{v_{24}}}}{2} + \log{\delimiterswithbreaks{(}{)}{- v_{35}}} - \log{\delimiterswithbreaks{(}{)}{- v_{45}}}}\nonumber \\
\Gamma^{(1)}_{2,4} =&\frac{N_{c}^{2} - 4}{2 N_{c}^{2}}\breaktowidth{0.8\linewidth}{\delimiterswithbreaks{[}{]}{\log{\delimiterswithbreaks{(}{)}{v_{13}}} - \log{\delimiterswithbreaks{(}{)}{v_{14}}} - \log{\delimiterswithbreaks{(}{)}{v_{23}}} + \log{\delimiterswithbreaks{(}{)}{v_{24}}}}}\nonumber \\
\Gamma^{(1)}_{3,1} =&\breaktowidth{0.8\linewidth}{- \log{\delimiterswithbreaks{(}{)}{v_{13}}} - \log{\delimiterswithbreaks{(}{)}{v_{14}}} + 2 \log{\delimiterswithbreaks{(}{)}{v_{15}}} + \log{\delimiterswithbreaks{(}{)}{v_{23}}} + \log{\delimiterswithbreaks{(}{)}{v_{24}}} - 2 \log{\delimiterswithbreaks{(}{)}{v_{25}}}}\nonumber \\
\Gamma^{(1)}_{3,2} =&\breaktowidth{0.8\linewidth}{- \log{\delimiterswithbreaks{(}{)}{v_{13}}} + \log{\delimiterswithbreaks{(}{)}{v_{14}}} - \log{\delimiterswithbreaks{(}{)}{v_{23}}} + \log{\delimiterswithbreaks{(}{)}{v_{24}}} + 2 \log{\delimiterswithbreaks{(}{)}{- v_{35}}} - 2 \log{\delimiterswithbreaks{(}{)}{- v_{45}}}}\nonumber \\
\Gamma^{(1)}_{3,3} =&\frac{1}{2 N_{c}}\breaktowidth{0.8\linewidth}{\delimiterswithbreaks{[}{]}{2 L_{\beta} + N_{c}^{2} \delimiterswithbreaks{(}{)}{- 2 \log{\delimiterswithbreaks{(}{)}{\nu_{5}}} + \log{\delimiterswithbreaks{(}{)}{v_{15}}} + \log{\delimiterswithbreaks{(}{)}{v_{25}}} + \log{\delimiterswithbreaks{(}{)}{- v_{35}}} + \log{\delimiterswithbreaks{(}{)}{- v_{45}}} + 2 \log{\delimiterswithbreaks{(}{)}{\frac{s}{m_{t}^{2}}}} - \log{\delimiterswithbreaks{(}{)}{16}} + 2} + \delimiterswithbreaks{(}{)}{N_{c}^{2} - 2} \delimiterswithbreaks{(}{)}{\log{\delimiterswithbreaks{(}{)}{v_{13}}} + \log{\delimiterswithbreaks{(}{)}{v_{24}}}} + \delimiterswithbreaks{(}{)}{N_{c}^{2} - 1} \delimiterswithbreaks{(}{)}{- \log{\delimiterswithbreaks{(}{)}{\nu_{1}}} - \log{\delimiterswithbreaks{(}{)}{\nu_{2}}}} + 2 \log{\delimiterswithbreaks{(}{)}{v_{14}}} + 2 \log{\delimiterswithbreaks{(}{)}{v_{23}}} + \log{\delimiterswithbreaks{(}{)}{4}} + 2 i \pi}}\nonumber \\
\Gamma^{(1)}_{3,4} =&\frac{N_{c}^{2} - 4}{2 N_{c}}\breaktowidth{0.8\linewidth}{\delimiterswithbreaks{[}{]}{- \log{\delimiterswithbreaks{(}{)}{v_{13}}} + \log{\delimiterswithbreaks{(}{)}{v_{15}}} + \log{\delimiterswithbreaks{(}{)}{v_{24}}} - \log{\delimiterswithbreaks{(}{)}{v_{25}}} + \log{\delimiterswithbreaks{(}{)}{- v_{35}}} - \log{\delimiterswithbreaks{(}{)}{- v_{45}}}}}\nonumber \\
\Gamma^{(1)}_{4,1} =&\breaktowidth{0.8\linewidth}{\log{\delimiterswithbreaks{(}{)}{v_{13}}} - \log{\delimiterswithbreaks{(}{)}{v_{14}}} - \log{\delimiterswithbreaks{(}{)}{v_{23}}} + \log{\delimiterswithbreaks{(}{)}{v_{24}}}}\nonumber \\
\Gamma^{(1)}_{4,2} =&\breaktowidth{0.8\linewidth}{\log{\delimiterswithbreaks{(}{)}{v_{13}}} - \log{\delimiterswithbreaks{(}{)}{v_{14}}} - \log{\delimiterswithbreaks{(}{)}{v_{23}}} + \log{\delimiterswithbreaks{(}{)}{v_{24}}}}\nonumber \\
\Gamma^{(1)}_{4,3} =&\frac{N_{c}}{2}\breaktowidth{0.8\linewidth}{\delimiterswithbreaks{[}{]}{- \log{\delimiterswithbreaks{(}{)}{v_{13}}} + \log{\delimiterswithbreaks{(}{)}{v_{15}}} + \log{\delimiterswithbreaks{(}{)}{v_{24}}} - \log{\delimiterswithbreaks{(}{)}{v_{25}}} + \log{\delimiterswithbreaks{(}{)}{- v_{35}}} - \log{\delimiterswithbreaks{(}{)}{- v_{45}}}}}\nonumber \\
\Gamma^{(1)}_{4,4} =&\frac{1}{2 N_{c}}\breaktowidth{0.8\linewidth}{\delimiterswithbreaks{[}{]}{2 L_{\beta} + N_{c}^{2} \delimiterswithbreaks{(}{)}{- 2 \log{\delimiterswithbreaks{(}{)}{\nu_{5}}} + \log{\delimiterswithbreaks{(}{)}{v_{15}}} + \log{\delimiterswithbreaks{(}{)}{v_{25}}} + \log{\delimiterswithbreaks{(}{)}{- v_{35}}} + \log{\delimiterswithbreaks{(}{)}{- v_{45}}} + 2 \log{\delimiterswithbreaks{(}{)}{\frac{s}{m_{t}^{2}}}} - \log{\delimiterswithbreaks{(}{)}{16}} + 2} + \delimiterswithbreaks{(}{)}{N_{c}^{2} - 6} \delimiterswithbreaks{(}{)}{\log{\delimiterswithbreaks{(}{)}{v_{13}}} + \log{\delimiterswithbreaks{(}{)}{v_{24}}}} + \delimiterswithbreaks{(}{)}{N_{c}^{2} - 1} \delimiterswithbreaks{(}{)}{- \log{\delimiterswithbreaks{(}{)}{\nu_{1}}} - \log{\delimiterswithbreaks{(}{)}{\nu_{2}}}} + 6 \log{\delimiterswithbreaks{(}{)}{v_{14}}} + 6 \log{\delimiterswithbreaks{(}{)}{v_{23}}} + \log{\delimiterswithbreaks{(}{)}{4}} + 2 i \pi}}\nonumber \\
\end{align*}

}}

In the previous formula, $L_\beta$ is the velocity-dependent eikonal function:
\begin{equation}
	L_{\beta} = \frac{1-2m_t^2/s_{t\bar t}}{\beta} \left(\ln \frac{1-\beta}{1+\beta} +\pi i \right)\, ,
\end{equation}
with $\beta = \sqrt{1-4m_t^2/s_{t\bar t}}$, where $s_{t\bar t}=(p_3+p_4)^2$ is the
squared invariant mass of the top-quark pair, $m_t$ the top-quark mass and
$v_{ij}=v_iv_j$, see Eq.~(\ref{eq:v_i}).
Logarithms of negative arguments are defined by $\log(-x)=\log(x)-i\pi$ for $x>0$, see
Section~\ref{sec:4softcalc}. They arise for the dipoles formed by the outgoing jet and
the top quark or antiquark. For the dipole of the two incoming partons, the resulting
imaginary part appears explicitly in the expressions above.
For the corrections with respect to earlier versions of this article, see the note
at the end of Section~\ref{sec:conclu}.
Those cases where $\Gamma^{AB} \ne \Gamma^{BA}$ are related to the fact that
the corresponding color components $\mathcal{F}_{ij}^{AB} \ne
\mathcal{F}_{ij}^{BA}$, i.e. the $\mathcal{F}_{ij}$ matrix is in general
non-symmetric.

\subsection{$gg$ channel}
\label{sec:ggchannel}

The results of our calculation of the Born-level soft matrix and one-loop color
factors, using the color basis taken from Ref.~\cite{Sjodahl:2008fz} and
reported in Appendix~\ref{apx:basis_gg} for completeness, with basis elements
ordered as listed, are in agreement with those published in
Ref.~\cite{Sjodahl:2008fz}, confirming the correctness of the latter. For
explicit expressions, we address the interested reader to that reference. On
the other hand, the complete result that we obtained for the one-loop components
of the soft anomalous dimension matrix including both the color and the
kinematic factors, reads as follows:

{\small{
\allowdisplaybreaks
\setlength{\jot}{10pt}
\begin{align*}
\Gamma^{(1)}_{1,1} =&\frac{1}{N_{c}}\breaktowidth{0.8\linewidth}{\delimiterswithbreaks{[}{]}{L_{\beta} + N_{c}^{2} \delimiterswithbreaks{(}{)}{- \log{\delimiterswithbreaks{(}{)}{\nu_{1}}} - \log{\delimiterswithbreaks{(}{)}{\nu_{2}}} - \log{\delimiterswithbreaks{(}{)}{\nu_{5}}} + \log{\delimiterswithbreaks{(}{)}{- v_{35}}} + \log{\delimiterswithbreaks{(}{)}{- v_{45}}} + \log{\delimiterswithbreaks{(}{)}{\frac{s}{m_{t}^{2}}}} - \log{\delimiterswithbreaks{(}{)}{8}} + 2 - 2 i \pi} + 1}}\nonumber \\
\Gamma^{(1)}_{1,2} =&\frac{2 N_{c}}{N_{c}^{2} - 1}\breaktowidth{0.8\linewidth}{\delimiterswithbreaks{[}{]}{\log{\delimiterswithbreaks{(}{)}{v_{13}}} - \log{\delimiterswithbreaks{(}{)}{v_{14}}} - \log{\delimiterswithbreaks{(}{)}{v_{23}}} + \log{\delimiterswithbreaks{(}{)}{v_{24}}}}}\nonumber \\
\Gamma^{(1)}_{1,3} =&0\nonumber \\
\Gamma^{(1)}_{1,4} =&\frac{N_{c}^{2}}{N_{c}^{2} - 1}\breaktowidth{0.8\linewidth}{\delimiterswithbreaks{[}{]}{- \log{\delimiterswithbreaks{(}{)}{v_{13}}} - \log{\delimiterswithbreaks{(}{)}{v_{14}}} + 2 \log{\delimiterswithbreaks{(}{)}{v_{15}}} + \log{\delimiterswithbreaks{(}{)}{v_{23}}} + \log{\delimiterswithbreaks{(}{)}{v_{24}}} - 2 \log{\delimiterswithbreaks{(}{)}{v_{25}}}}}\nonumber \\
\Gamma^{(1)}_{1,5} =&0\nonumber \\
\Gamma^{(1)}_{1,6} =&\frac{N_{c}^{2} - 4}{N_{c}^{2} - 1}\breaktowidth{0.8\linewidth}{\delimiterswithbreaks{[}{]}{\log{\delimiterswithbreaks{(}{)}{v_{13}}} - \log{\delimiterswithbreaks{(}{)}{v_{14}}} - \log{\delimiterswithbreaks{(}{)}{v_{23}}} + \log{\delimiterswithbreaks{(}{)}{v_{24}}}}}\nonumber \\
\Gamma^{(1)}_{1,7} =&0\nonumber \\
\Gamma^{(1)}_{1,8} =&0\nonumber \\
\Gamma^{(1)}_{1,9} =&0\nonumber \\
\Gamma^{(1)}_{1,10} =&0\nonumber \\
\Gamma^{(1)}_{1,11} =&0\nonumber \\
\Gamma^{(1)}_{2,1} =&\frac{1}{N_{c}}\breaktowidth{0.8\linewidth}{\delimiterswithbreaks{[}{]}{\log{\delimiterswithbreaks{(}{)}{v_{13}}} - \log{\delimiterswithbreaks{(}{)}{v_{14}}} - \log{\delimiterswithbreaks{(}{)}{v_{23}}} + \log{\delimiterswithbreaks{(}{)}{v_{24}}}}}\nonumber \\
\Gamma^{(1)}_{2,2} =&\frac{1}{N_{c}}\breaktowidth{0.8\linewidth}{\delimiterswithbreaks{[}{]}{- L_{\beta} \delimiterswithbreaks{(}{)}{N_{c}^{2} - 1} + N_{c}^{2} \delimiterswithbreaks{(}{)}{- \log{\delimiterswithbreaks{(}{)}{\nu_{1}}} - \log{\delimiterswithbreaks{(}{)}{\nu_{2}}} - \log{\delimiterswithbreaks{(}{)}{\nu_{5}}} + \log{\delimiterswithbreaks{(}{)}{v_{15}}} + \log{\delimiterswithbreaks{(}{)}{v_{25}}} - \log{\delimiterswithbreaks{(}{)}{8}} + 2 - i \pi} + 1}}\nonumber \\
\Gamma^{(1)}_{2,3} =&0\nonumber \\
\Gamma^{(1)}_{2,4} =&\breaktowidth{0.8\linewidth}{- \frac{\log{\delimiterswithbreaks{(}{)}{v_{13}}}}{2} + \frac{\log{\delimiterswithbreaks{(}{)}{v_{14}}}}{2} - \frac{\log{\delimiterswithbreaks{(}{)}{v_{23}}}}{2} + \frac{\log{\delimiterswithbreaks{(}{)}{v_{24}}}}{2} + \log{\delimiterswithbreaks{(}{)}{- v_{35}}} - \log{\delimiterswithbreaks{(}{)}{- v_{45}}}}\nonumber \\
\Gamma^{(1)}_{2,5} =&0\nonumber \\
\Gamma^{(1)}_{2,6} =&0\nonumber \\
\Gamma^{(1)}_{2,7} =&\frac{N_{c}^{2} - 4}{2 N_{c}^{2}}\breaktowidth{0.8\linewidth}{\delimiterswithbreaks{[}{]}{\log{\delimiterswithbreaks{(}{)}{v_{13}}} - \log{\delimiterswithbreaks{(}{)}{v_{14}}} - \log{\delimiterswithbreaks{(}{)}{v_{23}}} + \log{\delimiterswithbreaks{(}{)}{v_{24}}}}}\nonumber \\
\Gamma^{(1)}_{2,8} =&0\nonumber \\
\Gamma^{(1)}_{2,9} =&0\nonumber \\
\Gamma^{(1)}_{2,10} =&\frac{N_{c} + 3}{4 N_{c} + 4}\breaktowidth{0.8\linewidth}{\delimiterswithbreaks{[}{]}{\log{\delimiterswithbreaks{(}{)}{v_{13}}} - \log{\delimiterswithbreaks{(}{)}{v_{14}}} - \log{\delimiterswithbreaks{(}{)}{v_{23}}} + \log{\delimiterswithbreaks{(}{)}{v_{24}}}}}\nonumber \\
\Gamma^{(1)}_{2,11} =&\frac{N_{c} - 3}{4 N_{c} - 4}\breaktowidth{0.8\linewidth}{\delimiterswithbreaks{[}{]}{\log{\delimiterswithbreaks{(}{)}{v_{13}}} - \log{\delimiterswithbreaks{(}{)}{v_{14}}} - \log{\delimiterswithbreaks{(}{)}{v_{23}}} + \log{\delimiterswithbreaks{(}{)}{v_{24}}}}}\nonumber \\
\Gamma^{(1)}_{3,1} =&0\nonumber \\
\Gamma^{(1)}_{3,2} =&0\nonumber \\
\Gamma^{(1)}_{3,3} =&\frac{1}{N_{c}}\breaktowidth{0.8\linewidth}{\delimiterswithbreaks{[}{]}{- L_{\beta} \delimiterswithbreaks{(}{)}{N_{c}^{2} - 1} + N_{c}^{2} \delimiterswithbreaks{(}{)}{- \log{\delimiterswithbreaks{(}{)}{\nu_{1}}} - \log{\delimiterswithbreaks{(}{)}{\nu_{2}}} - \log{\delimiterswithbreaks{(}{)}{\nu_{5}}} + \log{\delimiterswithbreaks{(}{)}{v_{15}}} + \log{\delimiterswithbreaks{(}{)}{v_{25}}} - \log{\delimiterswithbreaks{(}{)}{8}} + 2 - i \pi} + 1}}\nonumber \\
\Gamma^{(1)}_{3,4} =&0\nonumber \\
\Gamma^{(1)}_{3,5} =&\breaktowidth{0.8\linewidth}{- \frac{\log{\delimiterswithbreaks{(}{)}{v_{13}}}}{2} + \frac{\log{\delimiterswithbreaks{(}{)}{v_{14}}}}{2} - \frac{\log{\delimiterswithbreaks{(}{)}{v_{23}}}}{2} + \frac{\log{\delimiterswithbreaks{(}{)}{v_{24}}}}{2} + \log{\delimiterswithbreaks{(}{)}{- v_{35}}} - \log{\delimiterswithbreaks{(}{)}{- v_{45}}}}\nonumber \\
\Gamma^{(1)}_{3,6} =&\breaktowidth{0.8\linewidth}{\frac{\log{\delimiterswithbreaks{(}{)}{v_{13}}}}{2} - \frac{\log{\delimiterswithbreaks{(}{)}{v_{14}}}}{2} - \frac{\log{\delimiterswithbreaks{(}{)}{v_{23}}}}{2} + \frac{\log{\delimiterswithbreaks{(}{)}{v_{24}}}}{2}}\nonumber \\
\Gamma^{(1)}_{3,7} =&0\nonumber \\
\Gamma^{(1)}_{3,8} =&0\nonumber \\
\Gamma^{(1)}_{3,9} =&\breaktowidth{0.8\linewidth}{\frac{\log{\delimiterswithbreaks{(}{)}{v_{13}}}}{2} - \frac{\log{\delimiterswithbreaks{(}{)}{v_{14}}}}{2} - \frac{\log{\delimiterswithbreaks{(}{)}{v_{23}}}}{2} + \frac{\log{\delimiterswithbreaks{(}{)}{v_{24}}}}{2}}\nonumber \\
\Gamma^{(1)}_{3,10} =&0\nonumber \\
\Gamma^{(1)}_{3,11} =&0\nonumber \\
\Gamma^{(1)}_{4,1} =&\breaktowidth{0.8\linewidth}{- \log{\delimiterswithbreaks{(}{)}{v_{13}}} - \log{\delimiterswithbreaks{(}{)}{v_{14}}} + 2 \log{\delimiterswithbreaks{(}{)}{v_{15}}} + \log{\delimiterswithbreaks{(}{)}{v_{23}}} + \log{\delimiterswithbreaks{(}{)}{v_{24}}} - 2 \log{\delimiterswithbreaks{(}{)}{v_{25}}}}\nonumber \\
\Gamma^{(1)}_{4,2} =&\breaktowidth{0.8\linewidth}{- \log{\delimiterswithbreaks{(}{)}{v_{13}}} + \log{\delimiterswithbreaks{(}{)}{v_{14}}} - \log{\delimiterswithbreaks{(}{)}{v_{23}}} + \log{\delimiterswithbreaks{(}{)}{v_{24}}} + 2 \log{\delimiterswithbreaks{(}{)}{- v_{35}}} - 2 \log{\delimiterswithbreaks{(}{)}{- v_{45}}}}\nonumber \\
\Gamma^{(1)}_{4,3} =&0\nonumber \\
\Gamma^{(1)}_{4,4} =&\frac{1}{N_{c}}\breaktowidth{0.8\linewidth}{\delimiterswithbreaks{[}{]}{L_{\beta} + N_{c}^{2} \delimiterswithbreaks{(}{)}{- \log{\delimiterswithbreaks{(}{)}{\nu_{1}}} - \log{\delimiterswithbreaks{(}{)}{\nu_{2}}} - \log{\delimiterswithbreaks{(}{)}{\nu_{5}}} + \frac{\log{\delimiterswithbreaks{(}{)}{v_{13}}}}{4} + \frac{\log{\delimiterswithbreaks{(}{)}{v_{14}}}}{4} + \frac{\log{\delimiterswithbreaks{(}{)}{v_{15}}}}{2} + \frac{\log{\delimiterswithbreaks{(}{)}{v_{23}}}}{4} + \frac{\log{\delimiterswithbreaks{(}{)}{v_{24}}}}{4} + \frac{\log{\delimiterswithbreaks{(}{)}{v_{25}}}}{2} + \frac{\log{\delimiterswithbreaks{(}{)}{- v_{35}}}}{2} + \frac{\log{\delimiterswithbreaks{(}{)}{- v_{45}}}}{2} + \log{\delimiterswithbreaks{(}{)}{\frac{s}{m_{t}^{2}}}} - \frac{\log{\delimiterswithbreaks{(}{)}{4096}}}{4} + 2 - i \pi} + 1}}\nonumber \\
\Gamma^{(1)}_{4,5} =&\frac{N_{c}^{2} - 4}{4 N_{c}}\breaktowidth{0.8\linewidth}{\delimiterswithbreaks{[}{]}{\log{\delimiterswithbreaks{(}{)}{v_{13}}} - \log{\delimiterswithbreaks{(}{)}{v_{14}}} - \log{\delimiterswithbreaks{(}{)}{v_{23}}} + \log{\delimiterswithbreaks{(}{)}{v_{24}}}}}\nonumber \\
\Gamma^{(1)}_{4,6} =&\frac{N_{c}^{2} - 4}{4 N_{c}}\breaktowidth{0.8\linewidth}{\delimiterswithbreaks{[}{]}{- \log{\delimiterswithbreaks{(}{)}{v_{13}}} + \log{\delimiterswithbreaks{(}{)}{v_{14}}} - \log{\delimiterswithbreaks{(}{)}{v_{23}}} + \log{\delimiterswithbreaks{(}{)}{v_{24}}} + 2 \log{\delimiterswithbreaks{(}{)}{- v_{35}}} - 2 \log{\delimiterswithbreaks{(}{)}{- v_{45}}}}}\nonumber \\
\Gamma^{(1)}_{4,7} =&\frac{N_{c}^{2} - 4}{4 N_{c}}\breaktowidth{0.8\linewidth}{\delimiterswithbreaks{[}{]}{- \log{\delimiterswithbreaks{(}{)}{v_{13}}} - \log{\delimiterswithbreaks{(}{)}{v_{14}}} + 2 \log{\delimiterswithbreaks{(}{)}{v_{15}}} + \log{\delimiterswithbreaks{(}{)}{v_{23}}} + \log{\delimiterswithbreaks{(}{)}{v_{24}}} - 2 \log{\delimiterswithbreaks{(}{)}{v_{25}}}}}\nonumber \\
\Gamma^{(1)}_{4,8} =&0\nonumber \\
\Gamma^{(1)}_{4,9} =&0\nonumber \\
\Gamma^{(1)}_{4,10} =&\frac{N_{c} + 3}{4 N_{c} + 4}\breaktowidth{0.8\linewidth}{\delimiterswithbreaks{[}{]}{\log{\delimiterswithbreaks{(}{)}{v_{13}}} + \log{\delimiterswithbreaks{(}{)}{v_{14}}} - 2 \log{\delimiterswithbreaks{(}{)}{v_{15}}} - \log{\delimiterswithbreaks{(}{)}{v_{23}}} - \log{\delimiterswithbreaks{(}{)}{v_{24}}} + 2 \log{\delimiterswithbreaks{(}{)}{v_{25}}}}}\nonumber \\
\Gamma^{(1)}_{4,11} =&\frac{N_{c} - 3}{4 N_{c} - 4}\breaktowidth{0.8\linewidth}{\delimiterswithbreaks{[}{]}{- \log{\delimiterswithbreaks{(}{)}{v_{13}}} - \log{\delimiterswithbreaks{(}{)}{v_{14}}} + 2 \log{\delimiterswithbreaks{(}{)}{v_{15}}} + \log{\delimiterswithbreaks{(}{)}{v_{23}}} + \log{\delimiterswithbreaks{(}{)}{v_{24}}} - 2 \log{\delimiterswithbreaks{(}{)}{v_{25}}}}}\nonumber \\
\Gamma^{(1)}_{5,1} =&0\nonumber \\
\Gamma^{(1)}_{5,2} =&0\nonumber \\
\Gamma^{(1)}_{5,3} =&\breaktowidth{0.8\linewidth}{- \log{\delimiterswithbreaks{(}{)}{v_{13}}} + \log{\delimiterswithbreaks{(}{)}{v_{14}}} - \log{\delimiterswithbreaks{(}{)}{v_{23}}} + \log{\delimiterswithbreaks{(}{)}{v_{24}}} + 2 \log{\delimiterswithbreaks{(}{)}{- v_{35}}} - 2 \log{\delimiterswithbreaks{(}{)}{- v_{45}}}}\nonumber \\
\Gamma^{(1)}_{5,4} =&\frac{N_{c}}{4}\breaktowidth{0.8\linewidth}{\delimiterswithbreaks{[}{]}{\log{\delimiterswithbreaks{(}{)}{v_{13}}} - \log{\delimiterswithbreaks{(}{)}{v_{14}}} - \log{\delimiterswithbreaks{(}{)}{v_{23}}} + \log{\delimiterswithbreaks{(}{)}{v_{24}}}}}\nonumber \\
\Gamma^{(1)}_{5,5} =&\frac{1}{N_{c}}\breaktowidth{0.8\linewidth}{\delimiterswithbreaks{[}{]}{L_{\beta} + N_{c}^{2} \delimiterswithbreaks{(}{)}{- \log{\delimiterswithbreaks{(}{)}{\nu_{1}}} - \log{\delimiterswithbreaks{(}{)}{\nu_{2}}} - \log{\delimiterswithbreaks{(}{)}{\nu_{5}}} + \frac{\log{\delimiterswithbreaks{(}{)}{v_{13}}}}{4} + \frac{\log{\delimiterswithbreaks{(}{)}{v_{14}}}}{4} + \frac{\log{\delimiterswithbreaks{(}{)}{v_{15}}}}{2} + \frac{\log{\delimiterswithbreaks{(}{)}{v_{23}}}}{4} + \frac{\log{\delimiterswithbreaks{(}{)}{v_{24}}}}{4} + \frac{\log{\delimiterswithbreaks{(}{)}{v_{25}}}}{2} + \frac{\log{\delimiterswithbreaks{(}{)}{- v_{35}}}}{2} + \frac{\log{\delimiterswithbreaks{(}{)}{- v_{45}}}}{2} + \log{\delimiterswithbreaks{(}{)}{\frac{s}{m_{t}^{2}}}} - \frac{\log{\delimiterswithbreaks{(}{)}{4096}}}{4} + 2 - i \pi} + 1}}\nonumber \\
\Gamma^{(1)}_{5,6} =&\frac{N_{c}}{4}\breaktowidth{0.8\linewidth}{\delimiterswithbreaks{[}{]}{- \log{\delimiterswithbreaks{(}{)}{v_{13}}} - \log{\delimiterswithbreaks{(}{)}{v_{14}}} + 2 \log{\delimiterswithbreaks{(}{)}{v_{15}}} + \log{\delimiterswithbreaks{(}{)}{v_{23}}} + \log{\delimiterswithbreaks{(}{)}{v_{24}}} - 2 \log{\delimiterswithbreaks{(}{)}{v_{25}}}}}\nonumber \\
\Gamma^{(1)}_{5,7} =&\frac{N_{c}^{2} - 4}{4 N_{c}}\breaktowidth{0.8\linewidth}{\delimiterswithbreaks{[}{]}{- \log{\delimiterswithbreaks{(}{)}{v_{13}}} + \log{\delimiterswithbreaks{(}{)}{v_{14}}} - \log{\delimiterswithbreaks{(}{)}{v_{23}}} + \log{\delimiterswithbreaks{(}{)}{v_{24}}} + 2 \log{\delimiterswithbreaks{(}{)}{- v_{35}}} - 2 \log{\delimiterswithbreaks{(}{)}{- v_{45}}}}}\nonumber \\
\Gamma^{(1)}_{5,8} =&\breaktowidth{0.8\linewidth}{\frac{\log{\delimiterswithbreaks{(}{)}{v_{13}}}}{2} - \frac{\log{\delimiterswithbreaks{(}{)}{v_{14}}}}{2} - \frac{\log{\delimiterswithbreaks{(}{)}{v_{23}}}}{2} + \frac{\log{\delimiterswithbreaks{(}{)}{v_{24}}}}{2}}\nonumber \\
\Gamma^{(1)}_{5,9} =&0\nonumber \\
\Gamma^{(1)}_{5,10} =&0\nonumber \\
\Gamma^{(1)}_{5,11} =&0\nonumber \\
\Gamma^{(1)}_{6,1} =&\breaktowidth{0.8\linewidth}{\log{\delimiterswithbreaks{(}{)}{v_{13}}} - \log{\delimiterswithbreaks{(}{)}{v_{14}}} - \log{\delimiterswithbreaks{(}{)}{v_{23}}} + \log{\delimiterswithbreaks{(}{)}{v_{24}}}}\nonumber \\
\Gamma^{(1)}_{6,2} =&0\nonumber \\
\Gamma^{(1)}_{6,3} =&\breaktowidth{0.8\linewidth}{\log{\delimiterswithbreaks{(}{)}{v_{13}}} - \log{\delimiterswithbreaks{(}{)}{v_{14}}} - \log{\delimiterswithbreaks{(}{)}{v_{23}}} + \log{\delimiterswithbreaks{(}{)}{v_{24}}}}\nonumber \\
\Gamma^{(1)}_{6,4} =&\frac{N_{c}}{4}\breaktowidth{0.8\linewidth}{\delimiterswithbreaks{[}{]}{- \log{\delimiterswithbreaks{(}{)}{v_{13}}} + \log{\delimiterswithbreaks{(}{)}{v_{14}}} - \log{\delimiterswithbreaks{(}{)}{v_{23}}} + \log{\delimiterswithbreaks{(}{)}{v_{24}}} + 2 \log{\delimiterswithbreaks{(}{)}{- v_{35}}} - 2 \log{\delimiterswithbreaks{(}{)}{- v_{45}}}}}\nonumber \\
\Gamma^{(1)}_{6,5} =&\frac{N_{c}}{4}\breaktowidth{0.8\linewidth}{\delimiterswithbreaks{[}{]}{- \log{\delimiterswithbreaks{(}{)}{v_{13}}} - \log{\delimiterswithbreaks{(}{)}{v_{14}}} + 2 \log{\delimiterswithbreaks{(}{)}{v_{15}}} + \log{\delimiterswithbreaks{(}{)}{v_{23}}} + \log{\delimiterswithbreaks{(}{)}{v_{24}}} - 2 \log{\delimiterswithbreaks{(}{)}{v_{25}}}}}\nonumber \\
\Gamma^{(1)}_{6,6} =&\frac{1}{N_{c}}\breaktowidth{0.8\linewidth}{\delimiterswithbreaks{[}{]}{L_{\beta} + N_{c}^{2} \delimiterswithbreaks{(}{)}{- \log{\delimiterswithbreaks{(}{)}{\nu_{1}}} - \log{\delimiterswithbreaks{(}{)}{\nu_{2}}} - \log{\delimiterswithbreaks{(}{)}{\nu_{5}}} + \frac{\log{\delimiterswithbreaks{(}{)}{v_{13}}}}{4} + \frac{\log{\delimiterswithbreaks{(}{)}{v_{14}}}}{4} + \frac{\log{\delimiterswithbreaks{(}{)}{v_{15}}}}{2} + \frac{\log{\delimiterswithbreaks{(}{)}{v_{23}}}}{4} + \frac{\log{\delimiterswithbreaks{(}{)}{v_{24}}}}{4} + \frac{\log{\delimiterswithbreaks{(}{)}{v_{25}}}}{2} + \frac{\log{\delimiterswithbreaks{(}{)}{- v_{35}}}}{2} + \frac{\log{\delimiterswithbreaks{(}{)}{- v_{45}}}}{2} + \log{\delimiterswithbreaks{(}{)}{\frac{s}{m_{t}^{2}}}} - \frac{\log{\delimiterswithbreaks{(}{)}{4096}}}{4} + 2 - i \pi} + 1}}\nonumber \\
\Gamma^{(1)}_{6,7} =&\frac{N_{c}^{2} - 12}{4 N_{c}}\breaktowidth{0.8\linewidth}{\delimiterswithbreaks{[}{]}{\log{\delimiterswithbreaks{(}{)}{v_{13}}} - \log{\delimiterswithbreaks{(}{)}{v_{14}}} - \log{\delimiterswithbreaks{(}{)}{v_{23}}} + \log{\delimiterswithbreaks{(}{)}{v_{24}}}}}\nonumber \\
\Gamma^{(1)}_{6,8} =&0\nonumber \\
\Gamma^{(1)}_{6,9} =&0\nonumber \\
\Gamma^{(1)}_{6,10} =&\frac{N_{c} (N_{c} + 3)}{4 N_{c}^{2} + 12 N_{c} + 8}\breaktowidth{0.8\linewidth}{\delimiterswithbreaks{[}{]}{\log{\delimiterswithbreaks{(}{)}{v_{13}}} - \log{\delimiterswithbreaks{(}{)}{v_{14}}} - \log{\delimiterswithbreaks{(}{)}{v_{23}}} + \log{\delimiterswithbreaks{(}{)}{v_{24}}}}}\nonumber \\
\Gamma^{(1)}_{6,11} =&\frac{N_{c} (N_{c} - 3)}{4 N_{c}^{2} - 12 N_{c} + 8}\breaktowidth{0.8\linewidth}{\delimiterswithbreaks{[}{]}{- \log{\delimiterswithbreaks{(}{)}{v_{13}}} + \log{\delimiterswithbreaks{(}{)}{v_{14}}} + \log{\delimiterswithbreaks{(}{)}{v_{23}}} - \log{\delimiterswithbreaks{(}{)}{v_{24}}}}}\nonumber \\
\Gamma^{(1)}_{7,1} =&0\nonumber \\
\Gamma^{(1)}_{7,2} =&\frac{N_{c}^{2}}{N_{c}^{2} - 4}\breaktowidth{0.8\linewidth}{\delimiterswithbreaks{[}{]}{\log{\delimiterswithbreaks{(}{)}{v_{13}}} - \log{\delimiterswithbreaks{(}{)}{v_{14}}} - \log{\delimiterswithbreaks{(}{)}{v_{23}}} + \log{\delimiterswithbreaks{(}{)}{v_{24}}}}}\nonumber \\
\Gamma^{(1)}_{7,3} =&0\nonumber \\
\Gamma^{(1)}_{7,4} =&\frac{N_{c}^{3}}{4 N_{c}^{2} - 16}\breaktowidth{0.8\linewidth}{\delimiterswithbreaks{[}{]}{- \log{\delimiterswithbreaks{(}{)}{v_{13}}} - \log{\delimiterswithbreaks{(}{)}{v_{14}}} + 2 \log{\delimiterswithbreaks{(}{)}{v_{15}}} + \log{\delimiterswithbreaks{(}{)}{v_{23}}} + \log{\delimiterswithbreaks{(}{)}{v_{24}}} - 2 \log{\delimiterswithbreaks{(}{)}{v_{25}}}}}\nonumber \\
\Gamma^{(1)}_{7,5} =&\frac{N_{c}}{4}\breaktowidth{0.8\linewidth}{\delimiterswithbreaks{[}{]}{- \log{\delimiterswithbreaks{(}{)}{v_{13}}} + \log{\delimiterswithbreaks{(}{)}{v_{14}}} - \log{\delimiterswithbreaks{(}{)}{v_{23}}} + \log{\delimiterswithbreaks{(}{)}{v_{24}}} + 2 \log{\delimiterswithbreaks{(}{)}{- v_{35}}} - 2 \log{\delimiterswithbreaks{(}{)}{- v_{45}}}}}\nonumber \\
\Gamma^{(1)}_{7,6} =&\frac{N_{c} (N_{c}^{2} - 12)}{4 N_{c}^{2} - 16}\breaktowidth{0.8\linewidth}{\delimiterswithbreaks{[}{]}{\log{\delimiterswithbreaks{(}{)}{v_{13}}} - \log{\delimiterswithbreaks{(}{)}{v_{14}}} - \log{\delimiterswithbreaks{(}{)}{v_{23}}} + \log{\delimiterswithbreaks{(}{)}{v_{24}}}}}\nonumber \\
\Gamma^{(1)}_{7,7} =&\frac{1}{N_{c}}\breaktowidth{0.8\linewidth}{\delimiterswithbreaks{[}{]}{L_{\beta} + N_{c}^{2} \delimiterswithbreaks{(}{)}{- \log{\delimiterswithbreaks{(}{)}{\nu_{1}}} - \log{\delimiterswithbreaks{(}{)}{\nu_{2}}} - \log{\delimiterswithbreaks{(}{)}{\nu_{5}}} + \frac{\log{\delimiterswithbreaks{(}{)}{v_{13}}}}{4} + \frac{\log{\delimiterswithbreaks{(}{)}{v_{14}}}}{4} + \frac{\log{\delimiterswithbreaks{(}{)}{v_{15}}}}{2} + \frac{\log{\delimiterswithbreaks{(}{)}{v_{23}}}}{4} + \frac{\log{\delimiterswithbreaks{(}{)}{v_{24}}}}{4} + \frac{\log{\delimiterswithbreaks{(}{)}{v_{25}}}}{2} + \frac{\log{\delimiterswithbreaks{(}{)}{- v_{35}}}}{2} + \frac{\log{\delimiterswithbreaks{(}{)}{- v_{45}}}}{2} + \log{\delimiterswithbreaks{(}{)}{\frac{s}{m_{t}^{2}}}} - \frac{\log{\delimiterswithbreaks{(}{)}{4096}}}{4} + 2 - i \pi} + 1}}\nonumber \\
\Gamma^{(1)}_{7,8} =&\frac{N_{c}^{2}}{2 N_{c}^{2} - 8}\breaktowidth{0.8\linewidth}{\delimiterswithbreaks{[}{]}{\log{\delimiterswithbreaks{(}{)}{v_{13}}} + \log{\delimiterswithbreaks{(}{)}{v_{14}}} - 2 \log{\delimiterswithbreaks{(}{)}{v_{15}}} - \log{\delimiterswithbreaks{(}{)}{v_{23}}} - \log{\delimiterswithbreaks{(}{)}{v_{24}}} + 2 \log{\delimiterswithbreaks{(}{)}{v_{25}}}}}\nonumber \\
\Gamma^{(1)}_{7,9} =&\frac{N_{c}}{N_{c}^{2} - 4}\breaktowidth{0.8\linewidth}{\delimiterswithbreaks{[}{]}{- \log{\delimiterswithbreaks{(}{)}{v_{13}}} + \log{\delimiterswithbreaks{(}{)}{v_{14}}} + \log{\delimiterswithbreaks{(}{)}{v_{23}}} - \log{\delimiterswithbreaks{(}{)}{v_{24}}}}}\nonumber \\
\Gamma^{(1)}_{7,10} =&0\nonumber \\
\Gamma^{(1)}_{7,11} =&0\nonumber \\
\Gamma^{(1)}_{8,1} =&0\nonumber \\
\Gamma^{(1)}_{8,2} =&0\nonumber \\
\Gamma^{(1)}_{8,3} =&0\nonumber \\
\Gamma^{(1)}_{8,4} =&0\nonumber \\
\Gamma^{(1)}_{8,5} =&\breaktowidth{0.8\linewidth}{\log{\delimiterswithbreaks{(}{)}{v_{13}}} - \log{\delimiterswithbreaks{(}{)}{v_{14}}} - \log{\delimiterswithbreaks{(}{)}{v_{23}}} + \log{\delimiterswithbreaks{(}{)}{v_{24}}}}\nonumber \\
\Gamma^{(1)}_{8,6} =&0\nonumber \\
\Gamma^{(1)}_{8,7} =&\breaktowidth{0.8\linewidth}{\log{\delimiterswithbreaks{(}{)}{v_{13}}} + \log{\delimiterswithbreaks{(}{)}{v_{14}}} - 2 \log{\delimiterswithbreaks{(}{)}{v_{15}}} - \log{\delimiterswithbreaks{(}{)}{v_{23}}} - \log{\delimiterswithbreaks{(}{)}{v_{24}}} + 2 \log{\delimiterswithbreaks{(}{)}{v_{25}}}}\nonumber \\
\Gamma^{(1)}_{8,8} =&\frac{1}{N_{c}}\breaktowidth{0.8\linewidth}{\delimiterswithbreaks{[}{]}{L_{\beta} + N_{c}^{2} \delimiterswithbreaks{(}{)}{- \log{\delimiterswithbreaks{(}{)}{\nu_{1}}} - \log{\delimiterswithbreaks{(}{)}{\nu_{2}}} - \log{\delimiterswithbreaks{(}{)}{\nu_{5}}} + \frac{\log{\delimiterswithbreaks{(}{)}{v_{13}}}}{2} + \frac{\log{\delimiterswithbreaks{(}{)}{v_{14}}}}{2} + \log{\delimiterswithbreaks{(}{)}{v_{15}}} + \frac{\log{\delimiterswithbreaks{(}{)}{v_{23}}}}{2} + \frac{\log{\delimiterswithbreaks{(}{)}{v_{24}}}}{2} + \log{\delimiterswithbreaks{(}{)}{v_{25}}} + \log{\delimiterswithbreaks{(}{)}{\frac{s}{m_{t}^{2}}}} - \frac{\log{\delimiterswithbreaks{(}{)}{64}}}{2} + 2} + 1}}\nonumber \\
\Gamma^{(1)}_{8,9} =&\breaktowidth{0.8\linewidth}{- \frac{\log{\delimiterswithbreaks{(}{)}{v_{13}}}}{2} + \frac{\log{\delimiterswithbreaks{(}{)}{v_{14}}}}{2} - \frac{\log{\delimiterswithbreaks{(}{)}{v_{23}}}}{2} + \frac{\log{\delimiterswithbreaks{(}{)}{v_{24}}}}{2} + \log{\delimiterswithbreaks{(}{)}{- v_{35}}} - \log{\delimiterswithbreaks{(}{)}{- v_{45}}}}\nonumber \\
\Gamma^{(1)}_{8,10} =&\frac{N_{c} (N_{c} + 3)}{4 N_{c} + 8}\breaktowidth{0.8\linewidth}{\delimiterswithbreaks{[}{]}{- \log{\delimiterswithbreaks{(}{)}{v_{13}}} - \log{\delimiterswithbreaks{(}{)}{v_{14}}} + 2 \log{\delimiterswithbreaks{(}{)}{v_{15}}} + \log{\delimiterswithbreaks{(}{)}{v_{23}}} + \log{\delimiterswithbreaks{(}{)}{v_{24}}} - 2 \log{\delimiterswithbreaks{(}{)}{v_{25}}}}}\nonumber \\
\Gamma^{(1)}_{8,11} =&\frac{N_{c} (N_{c} - 3)}{4 N_{c} - 8}\breaktowidth{0.8\linewidth}{\delimiterswithbreaks{[}{]}{\log{\delimiterswithbreaks{(}{)}{v_{13}}} + \log{\delimiterswithbreaks{(}{)}{v_{14}}} - 2 \log{\delimiterswithbreaks{(}{)}{v_{15}}} - \log{\delimiterswithbreaks{(}{)}{v_{23}}} - \log{\delimiterswithbreaks{(}{)}{v_{24}}} + 2 \log{\delimiterswithbreaks{(}{)}{v_{25}}}}}\nonumber \\
\Gamma^{(1)}_{9,1} =&0\nonumber \\
\Gamma^{(1)}_{9,2} =&0\nonumber \\
\Gamma^{(1)}_{9,3} =&\breaktowidth{0.8\linewidth}{2 \log{\delimiterswithbreaks{(}{)}{v_{13}}} - 2 \log{\delimiterswithbreaks{(}{)}{v_{14}}} - 2 \log{\delimiterswithbreaks{(}{)}{v_{23}}} + 2 \log{\delimiterswithbreaks{(}{)}{v_{24}}}}\nonumber \\
\Gamma^{(1)}_{9,4} =&0\nonumber \\
\Gamma^{(1)}_{9,5} =&0\nonumber \\
\Gamma^{(1)}_{9,6} =&0\nonumber \\
\Gamma^{(1)}_{9,7} =&\frac{1}{N_{c}}\breaktowidth{0.8\linewidth}{\delimiterswithbreaks{[}{]}{- 2 \log{\delimiterswithbreaks{(}{)}{v_{13}}} + 2 \log{\delimiterswithbreaks{(}{)}{v_{14}}} + 2 \log{\delimiterswithbreaks{(}{)}{v_{23}}} - 2 \log{\delimiterswithbreaks{(}{)}{v_{24}}}}}\nonumber \\
\Gamma^{(1)}_{9,8} =&\breaktowidth{0.8\linewidth}{- \frac{\log{\delimiterswithbreaks{(}{)}{v_{13}}}}{2} + \frac{\log{\delimiterswithbreaks{(}{)}{v_{14}}}}{2} - \frac{\log{\delimiterswithbreaks{(}{)}{v_{23}}}}{2} + \frac{\log{\delimiterswithbreaks{(}{)}{v_{24}}}}{2} + \log{\delimiterswithbreaks{(}{)}{- v_{35}}} - \log{\delimiterswithbreaks{(}{)}{- v_{45}}}}\nonumber \\
\Gamma^{(1)}_{9,9} =&\frac{1}{N_{c}}\breaktowidth{0.8\linewidth}{\delimiterswithbreaks{[}{]}{L_{\beta} + N_{c}^{2} \delimiterswithbreaks{(}{)}{- \log{\delimiterswithbreaks{(}{)}{\nu_{1}}} - \log{\delimiterswithbreaks{(}{)}{\nu_{2}}} - \log{\delimiterswithbreaks{(}{)}{\nu_{5}}} + \frac{\log{\delimiterswithbreaks{(}{)}{v_{13}}}}{2} + \frac{\log{\delimiterswithbreaks{(}{)}{v_{14}}}}{2} + \log{\delimiterswithbreaks{(}{)}{v_{15}}} + \frac{\log{\delimiterswithbreaks{(}{)}{v_{23}}}}{2} + \frac{\log{\delimiterswithbreaks{(}{)}{v_{24}}}}{2} + \log{\delimiterswithbreaks{(}{)}{v_{25}}} + \log{\delimiterswithbreaks{(}{)}{\frac{s}{m_{t}^{2}}}} - \frac{\log{\delimiterswithbreaks{(}{)}{64}}}{2} + 2} + 1}}\nonumber \\
\Gamma^{(1)}_{9,10} =&\frac{N_{c} (N_{c} + 3)}{4 N_{c} + 8}\breaktowidth{0.8\linewidth}{\delimiterswithbreaks{[}{]}{- \log{\delimiterswithbreaks{(}{)}{v_{13}}} + \log{\delimiterswithbreaks{(}{)}{v_{14}}} + \log{\delimiterswithbreaks{(}{)}{v_{23}}} - \log{\delimiterswithbreaks{(}{)}{v_{24}}}}}\nonumber \\
\Gamma^{(1)}_{9,11} =&\frac{N_{c} (N_{c} - 3)}{4 N_{c} - 8}\breaktowidth{0.8\linewidth}{\delimiterswithbreaks{[}{]}{- \log{\delimiterswithbreaks{(}{)}{v_{13}}} + \log{\delimiterswithbreaks{(}{)}{v_{14}}} + \log{\delimiterswithbreaks{(}{)}{v_{23}}} - \log{\delimiterswithbreaks{(}{)}{v_{24}}}}}\nonumber \\
\Gamma^{(1)}_{10,1} =&0\nonumber \\
\Gamma^{(1)}_{10,2} =&\breaktowidth{0.8\linewidth}{2 \log{\delimiterswithbreaks{(}{)}{v_{13}}} - 2 \log{\delimiterswithbreaks{(}{)}{v_{14}}} - 2 \log{\delimiterswithbreaks{(}{)}{v_{23}}} + 2 \log{\delimiterswithbreaks{(}{)}{v_{24}}}}\nonumber \\
\Gamma^{(1)}_{10,3} =&0\nonumber \\
\Gamma^{(1)}_{10,4} =&\breaktowidth{0.8\linewidth}{\log{\delimiterswithbreaks{(}{)}{v_{13}}} + \log{\delimiterswithbreaks{(}{)}{v_{14}}} - 2 \log{\delimiterswithbreaks{(}{)}{v_{15}}} - \log{\delimiterswithbreaks{(}{)}{v_{23}}} - \log{\delimiterswithbreaks{(}{)}{v_{24}}} + 2 \log{\delimiterswithbreaks{(}{)}{v_{25}}}}\nonumber \\
\Gamma^{(1)}_{10,5} =&0\nonumber \\
\Gamma^{(1)}_{10,6} =&\frac{N_{c} - 2}{N_{c}}\breaktowidth{0.8\linewidth}{\delimiterswithbreaks{[}{]}{\log{\delimiterswithbreaks{(}{)}{v_{13}}} - \log{\delimiterswithbreaks{(}{)}{v_{14}}} - \log{\delimiterswithbreaks{(}{)}{v_{23}}} + \log{\delimiterswithbreaks{(}{)}{v_{24}}}}}\nonumber \\
\Gamma^{(1)}_{10,7} =&0\nonumber \\
\Gamma^{(1)}_{10,8} =&\frac{- N_{c} \delimiterswithbreaks{(}{)}{N_{c} - 1} + 2}{2 N_{c}}\breaktowidth{0.8\linewidth}{\delimiterswithbreaks{[}{]}{\log{\delimiterswithbreaks{(}{)}{v_{13}}} + \log{\delimiterswithbreaks{(}{)}{v_{14}}} - 2 \log{\delimiterswithbreaks{(}{)}{v_{15}}} - \log{\delimiterswithbreaks{(}{)}{v_{23}}} - \log{\delimiterswithbreaks{(}{)}{v_{24}}} + 2 \log{\delimiterswithbreaks{(}{)}{v_{25}}}}}\nonumber \\
\Gamma^{(1)}_{10,9} =&\frac{- N_{c} \delimiterswithbreaks{(}{)}{N_{c} - 1} + 2}{2 N_{c}}\breaktowidth{0.8\linewidth}{\delimiterswithbreaks{[}{]}{\log{\delimiterswithbreaks{(}{)}{v_{13}}} - \log{\delimiterswithbreaks{(}{)}{v_{14}}} - \log{\delimiterswithbreaks{(}{)}{v_{23}}} + \log{\delimiterswithbreaks{(}{)}{v_{24}}}}}\nonumber \\
\Gamma^{(1)}_{10,10} =&\frac{1}{N_{c}}\breaktowidth{0.8\linewidth}{\delimiterswithbreaks{[}{]}{L_{\beta} + N_{c} \delimiterswithbreaks{(}{)}{N_{c} \delimiterswithbreaks{(}{)}{- \log{\delimiterswithbreaks{(}{)}{\nu_{1}}} - \log{\delimiterswithbreaks{(}{)}{\nu_{2}}} - \log{\delimiterswithbreaks{(}{)}{\nu_{5}}} + \log{\delimiterswithbreaks{(}{)}{\frac{s}{m_{t}^{2}}}} - \frac{\log{\delimiterswithbreaks{(}{)}{64}}}{2} + 2} + \delimiterswithbreaks{(}{)}{N_{c} + 1} \delimiterswithbreaks{(}{)}{\frac{\log{\delimiterswithbreaks{(}{)}{v_{13}}}}{2} + \frac{\log{\delimiterswithbreaks{(}{)}{v_{14}}}}{2} + \log{\delimiterswithbreaks{(}{)}{v_{15}}} + \frac{\log{\delimiterswithbreaks{(}{)}{v_{23}}}}{2} + \frac{\log{\delimiterswithbreaks{(}{)}{v_{24}}}}{2} + \log{\delimiterswithbreaks{(}{)}{v_{25}}}} - \log{\delimiterswithbreaks{(}{)}{- v_{35}}} - \log{\delimiterswithbreaks{(}{)}{- v_{45}}} + 2 i \pi} + 1}}\nonumber \\
\Gamma^{(1)}_{10,11} =&0\nonumber \\
\Gamma^{(1)}_{11,1} =&0\nonumber \\
\Gamma^{(1)}_{11,2} =&\breaktowidth{0.8\linewidth}{2 \log{\delimiterswithbreaks{(}{)}{v_{13}}} - 2 \log{\delimiterswithbreaks{(}{)}{v_{14}}} - 2 \log{\delimiterswithbreaks{(}{)}{v_{23}}} + 2 \log{\delimiterswithbreaks{(}{)}{v_{24}}}}\nonumber \\
\Gamma^{(1)}_{11,3} =&0\nonumber \\
\Gamma^{(1)}_{11,4} =&\breaktowidth{0.8\linewidth}{- \log{\delimiterswithbreaks{(}{)}{v_{13}}} - \log{\delimiterswithbreaks{(}{)}{v_{14}}} + 2 \log{\delimiterswithbreaks{(}{)}{v_{15}}} + \log{\delimiterswithbreaks{(}{)}{v_{23}}} + \log{\delimiterswithbreaks{(}{)}{v_{24}}} - 2 \log{\delimiterswithbreaks{(}{)}{v_{25}}}}\nonumber \\
\Gamma^{(1)}_{11,5} =&0\nonumber \\
\Gamma^{(1)}_{11,6} =&\frac{N_{c} + 2}{N_{c}}\breaktowidth{0.8\linewidth}{\delimiterswithbreaks{[}{]}{- \log{\delimiterswithbreaks{(}{)}{v_{13}}} + \log{\delimiterswithbreaks{(}{)}{v_{14}}} + \log{\delimiterswithbreaks{(}{)}{v_{23}}} - \log{\delimiterswithbreaks{(}{)}{v_{24}}}}}\nonumber \\
\Gamma^{(1)}_{11,7} =&0\nonumber \\
\Gamma^{(1)}_{11,8} =&\frac{N_{c} \delimiterswithbreaks{(}{)}{N_{c} + 1} - 2}{2 N_{c}}\breaktowidth{0.8\linewidth}{\delimiterswithbreaks{[}{]}{\log{\delimiterswithbreaks{(}{)}{v_{13}}} + \log{\delimiterswithbreaks{(}{)}{v_{14}}} - 2 \log{\delimiterswithbreaks{(}{)}{v_{15}}} - \log{\delimiterswithbreaks{(}{)}{v_{23}}} - \log{\delimiterswithbreaks{(}{)}{v_{24}}} + 2 \log{\delimiterswithbreaks{(}{)}{v_{25}}}}}\nonumber \\
\Gamma^{(1)}_{11,9} =&\frac{N_{c} \delimiterswithbreaks{(}{)}{N_{c} + 1} - 2}{2 N_{c}}\breaktowidth{0.8\linewidth}{\delimiterswithbreaks{[}{]}{- \log{\delimiterswithbreaks{(}{)}{v_{13}}} + \log{\delimiterswithbreaks{(}{)}{v_{14}}} + \log{\delimiterswithbreaks{(}{)}{v_{23}}} - \log{\delimiterswithbreaks{(}{)}{v_{24}}}}}\nonumber \\
\Gamma^{(1)}_{11,10} =&0\nonumber \\
\Gamma^{(1)}_{11,11} =&\frac{1}{N_{c}}\breaktowidth{0.8\linewidth}{\delimiterswithbreaks{[}{]}{L_{\beta} + N_{c} \delimiterswithbreaks{(}{)}{N_{c} \delimiterswithbreaks{(}{)}{- \log{\delimiterswithbreaks{(}{)}{\nu_{1}}} - \log{\delimiterswithbreaks{(}{)}{\nu_{2}}} - \log{\delimiterswithbreaks{(}{)}{\nu_{5}}} + \log{\delimiterswithbreaks{(}{)}{\frac{s}{m_{t}^{2}}}} - \frac{\log{\delimiterswithbreaks{(}{)}{64}}}{2} + 2} + \delimiterswithbreaks{(}{)}{N_{c} - 1} \delimiterswithbreaks{(}{)}{\frac{\log{\delimiterswithbreaks{(}{)}{v_{13}}}}{2} + \frac{\log{\delimiterswithbreaks{(}{)}{v_{14}}}}{2} + \log{\delimiterswithbreaks{(}{)}{v_{15}}} + \frac{\log{\delimiterswithbreaks{(}{)}{v_{23}}}}{2} + \frac{\log{\delimiterswithbreaks{(}{)}{v_{24}}}}{2} + \log{\delimiterswithbreaks{(}{)}{v_{25}}}} + \log{\delimiterswithbreaks{(}{)}{- v_{35}}} + \log{\delimiterswithbreaks{(}{)}{- v_{45}}} - 2 i \pi} + 1}}\nonumber \\
\end{align*}
}}
Notice that all ($A$, $B$) combinations for which $\Gamma^{AB} = 0$ are
characterized by $\mathcal{F}_{ij}^{AB} = 0$ for all ($i$,$j$) pairs. Also
notice that, whenever $\Gamma^{AB} = 0$, even $\Gamma^{BA} = 0$, because the
zero terms are disposed symmetrically in the non-symmetric matrix
$\mathcal{F}_{ij}$, i.e.,
if $\mathcal{F}_{ij}^{AB} = 0$, even $\mathcal{F}_{ij}^{BA} = 0$.

\subsection{Discussion in the light of existing literature}
We note that results of soft anomalous dimension calculation for $t\bar{t}j$
hadroproduction have been previously published in Ref.~\cite{Szarek:2018skl},
working with different color bases. Different choices of color bases lead to
different normalizations of the leading-order soft function matrix used in the
calculation of the soft anomalous dimension matrix according to
Eq.~(\ref{eq:get_gamma}).
Hence, an additional transformation must be applied to the results presented in
this work to be able to compare them with those of Ref.~\cite{Szarek:2018skl}.
Additionally, while our results refer to the ge\-ne\-ral axial gauge
($n\cdot A=0$), those of Ref.~\cite{Szarek:2018skl} are obtained in a specific
axial gauge (Weyl gauge, corresponding to $A_0=0$).
Accounting for these aspects in the comparison
allows to identify some differences with the published results of Ref.~\cite{Szarek:2018skl}.
In particular, in both the $gg \rightarrow t\bar{t}g$ and $q\bar{q} \rightarrow t\bar{t}g$ anomalous
dimension matrices some of the elements of the soft anomalous dimension
matrices turned out to be mixed up, i.e., they do not correspond to the chosen
color basis.
Additionally, for massless partons Ref.~\cite{Szarek:2018skl} uses the
following substitution $\nu_i=\frac{1}{2}$ for the gauge-dependent terms,
which is only true in the Weyl gauge in the center--of--mass frame of a
$2\rightarrow 2$ process.
As a consequence of these issues, the soft anomalous dimension matrices
presented in Ref.~\cite{Szarek:2018skl} are not correct.
On the other hand, the results reported in Subsections~\ref{sec:qqchannel}
and~\ref{sec:ggchannel} satisfy the consistency checks discussed in the
following, besides the check in Eq.~(\ref{eq:check0}).

\section{Probing the infrared  poles at NLO}
\label{sec:poles}

With the soft anomalous dimension matrices at hand, it is possible to calculate
the infrared (IR) poles of the virtual amplitudes of the parton-parton
$\rightarrow t\bar{t}j$ production process at NLO and compare them against the
IR pole structure obtained using the Catani-Seymour subtraction
formalism~\cite{Catani:1996vz,Catani:2002hc}.

In the Catani-Seymour formalism the expression for the NLO part of the NLO
cross section reads
\begin{equation}
\label{eq:master}
  \delta \sigma_{\NLO} =
   \int_{n+1} [ (d\sigma_{\real})_{\epsilon=0} + (dA)_{\epsilon=0} ]
  + \int_n [ d\sigma_{\virt} + \int_1 dA' ]_{\epsilon=0}
  + \int dx \int_n [ d\sigma_{\fac}+dA'' ]_{\epsilon=0}
  \, ,
\end{equation}
with
\begin{equation}
  0 =
  \int_{n+1} dA
 + \int_n \int_1 dA'
 + \int dx \int_n dA''
 \, .
\end{equation}
The terms $dA$, $dA'$ and $dA''$ are constructed in such a way to make
the individual integrands in Eq.~(\ref{eq:master}) finite.
$dA$ represents the sum of all dipoles,
$\int_1 dA'$ corresponds to the integral of the sum of all dipoles  over
the phase space of the additional soft and/or collinear parton causing
divergences, i.e. the so-called
${\mbox{\bf I}}$-term, and the term $dA''$ arises from mass factorization and
includes the so-called ${\mbox{\bf P}}$-, ${\mbox{\bf K}}$- and ${\mbox{\bf
H}}$-terms.

All these terms are built using information about the soft and
collinear factorization properties of QCD amplitudes.
The subtraction term $dA$ can be expressed as a sum of individual dipoles:
\begin{equation}
  \label{eq:master1}
  dA = \sum \Dipole(i,j;k) \, .
\end{equation}
This sum runs over all colored external partons in the scattering process.
The possible $\{i,j;k\}$ combinations are obtained from the
real corrections for which the subtraction term $dA$ is constructed.

As an example let us consider the expression for $\Dipole_{ij,k}(p_1,p_2,\dotsc, p_{n+1})$:
\begin{eqnarray}
  \label{eq:dipggk}
{\lefteqn{
  \Dipole_{ij,k}(p_1,p_2,\dotsc, p_{n+1}) \,=\,}}
\nonumber\\
&=& -{1\over {2 p_i \cdot p_j}} \,\,
  _n\bra 1,\ldots \widetilde{ij},\ldots,\tilde k,\ldots,n+1 |
  {\T{k}\cdot\T{ij}\over \T{ij}^2} V_{ij,k}
  | 1,\ldots \tilde{ij},\ldots,\tilde k,\ldots,n+1\ket_n
  \, ,
\end{eqnarray}
which describes the configuration
corresponding to the splitting of a final-state parton $\tilde{ij}$ (emitter)
into the partons $i$ and $j$ in the presence of a final-state
spectator parton $\tilde{k}$.
The $n$-parton matrix element is obtained from the original $n+1$-parton matrix
element by replacing $i$ and $j$ with a single parton $\tilde{ij}$ and $k$
with $\tilde{k}$.
In the previous equation, the color charge operator $\T{k}$
is associated with the emission of a gluon from the
parton $k$.
If the emitted gluon has color index $c$, the color-charge operator is defined as
\begin{equation}
	\T{k}=\mathcal{T}_k^c|c\ket
\end{equation}
and its action onto the color space is defined by
\begin{equation}
	\bra c_1,\ldots,c_k,\ldots c_m,c|\T{k}|b_1,\ldots,b_k,\ldots b_m\ket=
	\delta_{c_1b_1}\dotsc \mathcal{T}_{c_kb_k}^c\ldots\delta_{c_mb_m},
\end{equation}
where $\mathcal{T}_{cb}^a\equiv if_{cab}$ if the emitting parton $k$ is a gluon
(adjoint representation of the $\SU(\Nc)$),
$\mathcal{T}_{cb}^a\equiv T_{cb}^a$ if the emitting parton is a quark (fundamental
representation) and $\mathcal{T}_{cb}^a\equiv \overline T_{cb}^a=-T^{a}_{bc}$
in case of an emitting antiquark\footnote{\label{ft:catani}Notice the difference in the color factor
	convention of the Catani-Seymour formalism with respect to the eikonal
	Feynman rules, as well as to the standard ones, where both quarks and
	antiquarks get color factors with a plus sign.
	This might seem inconsistent at a first glance, but actually it is not,
	keeping in mind that in case of the eikonal rules, the kinematic factor of
	the antiquark gets an extra minus sign with respect to the quark one.
	From the point of view of the sign, the factorization into a color and
	a kinematic factor is not unique and the minus sign can be assigned to
	either part.
	When using the results of calculation of soft anomalous dimension matrices
	in the dipole formalism through Eq.~(\ref{eq:clbs_soft}), it is important
	to remember the aforementioned aspects regarding the sign conventions.}.
These operators satisfy the usual color-algebra relations:
\begin{align}
	\T{k}\cdot\T{j}&=\T{j}\cdot\T{k}\qquad \text{if}\quad k\neq j \, ,\\
	\T{k}^2 &= C_k\qquad\qquad\quad \forall \, k \, ,
\end{align}
where $C_k = C_A$ for gluons and $C_k = C_F$ for quarks and antiquarks.

The matrix element
\begin{eqnarray}
\label{eq:clbs}
_n\bra 1,\ldots \widetilde{ij},\ldots,\tilde k,\ldots,n+1 |
  \T{k}\cdot\T{ij}
  | 1,\ldots \tilde{ij},\ldots,\tilde k,\ldots,n+1\ket_n
\end{eqnarray}
is referred to in the literature as color linked Born amplitude squared (CLBS).
The CLBS uses as input the reduced kinematics.
The original momenta $(p_{i},p_{j},p_{k})$ are reduced to
$(\tilde{p}_{ij}, \tilde{p}_{k})$.
In the dipole subtraction formalism the reduced kinematics obey the on-shell
conditions and momentum conservation.
This makes it possible to evaluate the CLBS
analytically in a
straightforward way or~numerically by applying the existing codes for
tree-level calculations.
Notice that the color operators $\T{k}$~$\cdot$~$\T{ij}$~which enter the CLBS
calculation, after bracketing, produce nothing else but the color factors
also emerging from the calculation of the soft anomalous
dimension matrices.
Thus, the CLBS can be calculated as:
\begin{eqnarray}
\label{eq:clbs_soft}
_n\bra 1,\ldots \widetilde{ij},\ldots,\tilde k,\ldots,n+1 |
  \T{k}\cdot\T{ij}
  | 1,\ldots \tilde{ij},\ldots,\tilde k,\ldots,n+1\ket_n = \tr(H_{AB}^{(0)}\tilde\Gamma_{BC}^{(1)}),
\end{eqnarray}
where $H^{(0)}_{AB}$ are the components of the color decomposed Born-level
hard-scattering amplitude and $\tilde\Gamma^{(1)}_{BC}$ are the color factors
of the one loop soft anomalous dimension.
$\tilde\Gamma^{(1)}_{BC}$ may differ from the color factor in the expression
of $\Gamma^{(1)}_{BC}$ at most by an overall sign, cf.~Footnote~\ref{ft:catani}.

Using this formalism we evaluate the ${\mbox{\bf I}}$-terms, which require the
CLBS as input.
The calculation of ${\mbox{\bf I}}$, that was repeated for both the $q\bar{q}$-
and the $gg$-induced subprocesses,
was performed numerically using the automated tool {\tt
AutoDipole}~\cite{Hasegawa:2009tx}.
The results obtained were then compared with ${\mbox{\bf I}}$-terms calculated
analytically using Eq.~(\ref{eq:clbs_soft}), where we inserted the
$\tilde\Gamma^{(1)}$ matrix elements which are part of the $\Gamma^{(1)}$ ones
presented in this work.

Additionally we verified that the matrix element of the ${\mbox{\bf I}}$
operator, calculated making use of Eq.~(\ref{eq:clbs_soft}), satisfies the
following equation
\begin{equation}
_n\bra 1,\ldots \widetilde{ij},\ldots,\tilde k,\ldots,n+1 |
 \, {\mbox{\bf I}}\,  | 1,\ldots \tilde{ij},\ldots,\tilde k,\ldots,n+1\ket_n  =
  \frac{C_{-2}}{\epsilon^2} + \frac{C_{-1}}{\epsilon} + C_0
  \end{equation}
producing the expected structure in terms of finite terms, simple and double
poles and their coefficients.

All these calculations were  automatized using {\tt FORM} and the {\tt
SymPy}~\cite{sympy} library in the {\tt python} module {\tt
pyDipole}~\cite{pydipole}.
This way we could confirm the correctness of all color factors of the soft
anomalous dimension matrices and the validity of the color decomposition
procedure.

\section{Conclusions}
\label{sec:conclu}

Analytical expressions for all terms of the soft anomalous dimension matrices
at one-loop for $t\bar{t}j$ production in parton-parton scattering close to
threshold have been presented.
This allows for the calculation of the one-loop soft functions, essential
ingredients of the threshold factorization formula for the resummation of the
logarithms associated to soft gluon emission at NLL accuracy.
The expressions are provided in the most general axial gauge and have been
subject to a number of analytical and numerical checks.

When the program of computing threshold resummation effects for the
$t\bar{t}j$ hadroproduction process at NLO+NLL accuracy in QCD will be completed,
indirect determinations of the top-quark mass with improved accuracy will indeed become possible,
thus overcoming the limitations of previous extractions which have
relied on NLO (or NLO+PS) differential distributions~\cite{Alioli:2022lqo}.
In particular, with the availability of results incorporating threshold
resummation effects, a much wider $\rho_s$ interval will be accessible
in the normalized $\rho_s$ distribution with respect to the restricted one used so far.
This increases the robustness of the extraction procedure and reduces
the theoretical uncertainties on the extracted top-quark mass value.
It is a crucial advancement of QCD theory in view of the high-statistics data
foreseen for the High-Luminosity LHC phase and of the
increase of competitiveness of indirect top-quark mass extractions with respect
to the direct ones.

{\tt FORM} files with the results for immediate use in resummation
frameworks are also provided in ancillary files.

\paragraph{Note added in version 3.}
In earlier versions of this article the matrices of Section~\ref{sec:softdim}
contained three errors in their kinematic factors. They were found by comparing
the matrices with the universal structure of the infrared singularities of amplitudes
with massive partons~\cite{Becher:2009kw,Ferroglia:2009ep}, and they have the following
origin.
(i) The function $L_\beta$ had been taken over from the $2\to2$ results of
Ref.~\cite{Kidonakis:1997gm}, where $s$ is at the same time the partonic center-of-mass
energy squared and the invariant mass squared of the heavy-quark pair. For $t\bar tj$
production the correct variable is $s_{t\bar t}$.
(ii) The FORM calculation kept the logarithm of the dipole of the two incoming partons
as the unevaluated symbol $\ln(-\frac{1}{2})$. In the conversion of the FORM results to
the printed expressions this symbol was evaluated by computer algebra on the principal
branch, as $\ln\frac{1}{2}+i\pi$. The $i\epsilon$ prescription requires
$\ln\frac{1}{2}-i\pi$, as in the explicit results of Ref.~\cite{Kidonakis:1997gm}.
(iii) The massive-massless integral $I_1$ of Ref.~\cite{Kidonakis:1997gm}, which is
real for the configuration considered there, had been applied unchanged to the dipoles
formed by the outgoing jet and the outgoing top quark or antiquark. Their imaginary parts
were therefore missing. The general form given in Section~\ref{sec:4softcalc} replaces
$\log(v_{35})$ and $\log(v_{45})$ by $\log(-v_{35})$ and $\log(-v_{45})$,
that is, it adds $2i\pi\,(\mathbf{T}_3\cdot\mathbf{T}_5+\mathbf{T}_4\cdot\mathbf{T}_5)$
to $\Gamma^{(1)}$.
With these changes the matrices of Section~\ref{sec:softdim} agree with the universal
result, up to a real multiple of the unit matrix which contains the gauge-dependent terms,
for $N_c=3$ and $N_c=4$ at random phase-space points.
The color factors, the real parts of all logarithms other than $L_\beta$, and hence the
checks of Section~\ref{sec:poles}, are not affected. In the ancillary files the
errors (i) and (ii) concern only the meaning of the symbols \texttt{[L\_beta]} and
\texttt{[ln(-1/2)]}, which is now documented in \texttt{loader.frm}. The same file
applies the correction (iii).

\Acknowledgements
The work of B.C. was partially supported by the DAAD (German Academic Exchange Program) scholarship 57440925.
The work of M.V.G. and S.M. was supported in part by the Bundesministerium f\"ur Bildung und Forschung under contract 05H21GUCCA.
\\
\\

\appendix
\section{Color bases}
\label{apx:bases}
We use the color bases of Ref.~\cite{Sjodahl:2008fz}.
For completeness we list them below, with components in the same order in which
they are used in the rest of our manuscript.
In the following expressions for the basis elements, \{$T^e$\} are the
generators of the $\SU(\Nc)$ group in the fundamental representation.  They are
$N_c \times N_c$ matrices. $f^{emn}$ and $d^{emn}$ are the totally
antisymmetric and totally symmetric structure constants, obeying
\begin{eqnarray}
	f^{emn}&=2i\mathrm{Tr}([T^e, T^m]T^n) \, , \label{eq:f_abc}\\
	d^{emn}&=2\mathrm{Tr}(\{T^e,T^m\}T^n) \, , \label{eq:d_abc}
\end{eqnarray}
respectively.
The color indices $e$, $m$, $n$ run from 1 to $N_c^2 -1$.
\subsection{Color basis for $q_a\bar q_b \rightarrow q_c\bar q_dg_e$}
\label{apx:basis_qq}
\begin{eqnarray}
	(\Col_1)_{abcde}&=& \delta_{ab}T^e{}_{cd} \, , \nonumber \\
	(\Col_2)_{abcde}&=& \delta_{cd}T^e{}_{ba} \, , \nonumber \\
	(\Col_3)_{abcde}&=& T^{m}{}_{ba}T^{n}{}_{cd} i f_{m n e} \, , \nonumber \\
	(\Col_4)_{abcde}&=& T^{m}{}_{ba}T^{n}{}_{cd} d_{m n e} \, ,
\label{eq:qqbarqqbargBasis}
\end{eqnarray}
where the $a$, $b$, $c$, $d$ color indices run from 1 to $N_c$ and $e$, $m$, $n$
run from 1 to $N_c^2-1$.

\subsection{Color basis for $g_a g_b \rightarrow q_c\bar q_dg_e$}
\label{apx:basis_gg}
\begin{eqnarray}
(\Col_1)_{abcde}&=&  T^e {}_{cd} \delta _{ ab } \, , \nonumber\\
(\Col_2)_{abcde}&=& i f_{ abe } \delta _{ cd } \, , \nonumber\\
(\Col_3)_{abcde}&=& d_{ abe } \delta _{cd} \, , \nonumber\\
(\Col_4)_{abcde}&=& if_{ abn } if_{ me n }
    T^{m} {}_{cd} \, , \nonumber\\
(\Col_5)_{abcde}&=& d_{ ab n } if_{ m e n }
    T^{m} {}_{cd} \, , \nonumber\\
(\Col_6)_{abcde}&=& if_{ ab n } d_{ m e n }
    T^{m} {}_{cd} \, , \nonumber\\
(\Col_7)_{abcde}&=& d_{ a b n } d_{ m e n }
    T^{m} {}_{cd} \, , \nonumber\\
(\Col_8)_{abcde}&=& P^{10+\overline{10}}_{abme}  T^{m} {}_{cd} \, , \nonumber\\
(\Col_9)_{abcde}&=& P^{10-\overline{10}}_{abme} T^{m} {}_{cd} \, , \nonumber\\
    (\Col_{10})_{abcde}&=&-P^{27}_{abme} T^{m} {}_{cd} \, , \nonumber\\
    (\Col_{11})_{abcde}&=& P^{0}_{abme}T^{m} {}_{cd} \, ,
\label{eq:ggqqbargBasis}
\end{eqnarray}
where the $c$, $d$ color indices run from 1 to $N_c$, the $a$, $b$, $e$, $m$,
$n$ color indices run from 1 to $N_c^2-1$, and the multiplet projectors are
given as:
\begin{eqnarray}
   P^{0}_{abcd}&=&-\frac{N d_{{ab g}} d_{{cdg}}}{4(N-2)}-\frac{1}{2} f_{{ad g}}
   f_{{cbg}}+\frac{1}{4} f_{{abg}}
   f_{{cdg}} \nonumber\\
   & &+\frac{1}{4}\delta _{{ac}} \delta
   _{{bd}}+\frac{1}{4} \delta _{{ad}} \delta _{{cb}}
   -\frac{\delta _{{ab}} \delta _{{cd}}}{2(N-1)} \, , \nonumber\\
      P^{10-\overline{10}}_{abcd}&=&\frac{1}{2} d_{{acg}} i f_{{bgd}}-\frac{1}{2} d_{{bgd}}i f_{{acg}} \, , \nonumber\\
      P^{10+\overline{10}}_{abcd}&=&\frac{1}{2} \left(\delta _{{ac}} \delta _{{bd}}-\delta _{{ad}}
   \delta _{{cb}}\right)-\frac{f_{{abg}} f_{{cdg}}}{N} \, , \nonumber\\
   P^{27}_{abcd}&=&\frac{N d_{{abg}} d_{{cdg}}}{4(N+2)}+\frac{1}{2} f_{{adg}}
   f_{{cbg}}-\frac{1}{4} f_{{ab g}}
   f_{{cdg}}+\frac{1}{4} \delta _{{ad}} \delta
   _{{bc}}\nonumber\\
   & &+\frac{1}{4} \delta _{{ac}} \delta
   _{{bd}}+\frac{\delta _{{ab}} \delta _{{cd}}}{2 (N+1)}.
  \label{eq:ggBasis}
\end{eqnarray}

\section{A simple example of color decomposition}
\label{apx:col_dec}
\begin{figure}[h]
	\centering
	\includegraphics[width=0.8\columnwidth]{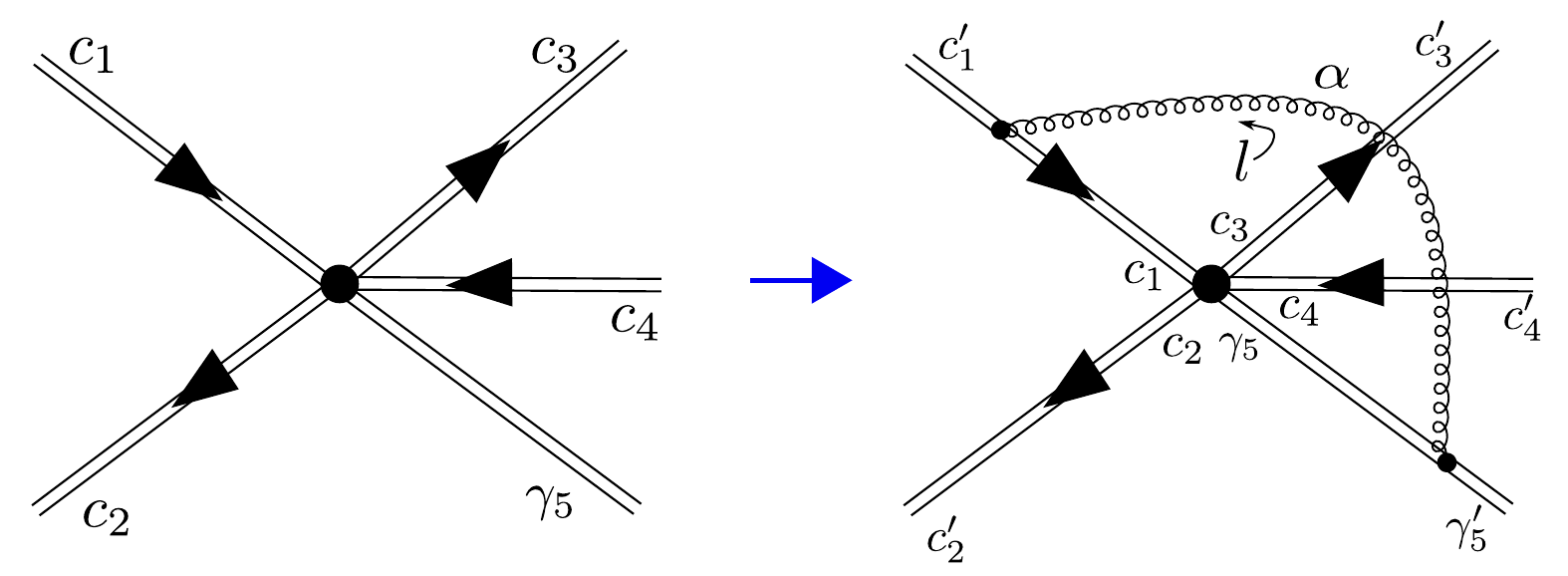}
	\caption{Example of color flow before and after soft gluon exchange.
	}
	\label{fig:decomp_example}
\end{figure}

In this Appendix we show how the color decomposition can be done in the
simplest cases, using as example the particular vertex correction graph
depicted on the right-hand side of Fig.~\ref{fig:decomp_example} among all
those for the $q\bar{q} \rightarrow q\bar{q} g$ subprocess
shown in Fig.~\ref{fig:vx_corr}.

Following the argumentation of Ref.~\cite{Sjodahl:2008fz}, one can use as
Born-level color basis $\{\Col_i\}$ the one in Appendix A.1, specifying the
color indices to the case of the graph at hand (see
Fig.~\ref{fig:decomp_example}):
\begin{equation}
	\renewcommand*{\arraystretch}{1.2}
	\begin{array}{l}
		\Col_1 = \delta_{c_1c_2}T^{\gamma_5}_{c_3c_4} \, ,\\
		\Col_2 = \delta_{c_3c_4}T^{\gamma_5}_{c_2c_1} \, ,\\
		\Col_3 = T^{\alpha}_{c_2c_1}T^{\beta}_{c_3c_4}if^{\alpha\beta\gamma_5} \, ,\\
		\Col_4 = T^{\alpha}_{c_2c_1}T^{\beta}_{c_3c_4}d^{\alpha\beta\gamma_5} \, ,
	\end{array}
	\label{eq:born_bas_qq}
\end{equation}
with $\alpha$, $\beta$, $\gamma_5$ running from 1 to $N_c^2-1$ and
$c_1$, $c_2$, $c_3$, $c_4$ running from 1 to $N_c$.
As shown in Fig.~\ref{fig:decomp_example},
the colors of two partons might change after a gluon exchange
(e.g., for the case at hand, $c_1 \rightarrow c_1^\prime$, $\gamma_5
\rightarrow \gamma_5^\prime$).
The basis in terms of color indices after gluon exchange $\{c_i^\prime\}$ is
then given by
\begin{equation}
	\renewcommand*{\arraystretch}{1.2}
	\begin{array}{l}
		\Col_1' = \delta_{c_1'c_2'}T^{\gamma_5'}_{c_3'c_4'} \, ,\\
		\Col_2' = \delta_{c_3'c_4'}T^{\gamma_5}_{c_2'c_1'} \, , \\
		\Col_3' = T^{\alpha}_{c_2'c_1'}T^{\beta}_{c_3'c_4'}if^{\alpha\beta\gamma_5'} \, ,\\
		\Col_4' = T^{\alpha}_{c_2'c_1'}T^{\beta}_{c_3'c_4'}d^{\alpha\beta\gamma_5'} \, .
	\end{array}
	\label{eq:born_bas_mod}
\end{equation}

By performing color tensor manipulations, making use of the Fierz identity for
the generators in the fundamental
representation:
\begin{equation}
	T^{\alpha}_{ij}T^{\alpha}_{kl}=\frac{1}{2}\left(\delta_{il}\delta_{jk}-\frac{1}{N_c}\delta_{ij}\delta_{kl}\right),
\end{equation}
and of the properties Eqs.~(\ref{eq:f_abc}) and (\ref{eq:d_abc})
for the structure constants, one can show that the vertex correction leads to
the following relation between the basis in terms of the Born-level color
indices and the basis in terms of color indices after gluon exchange:
\begin{align}
	\Col_1\mathscr{F}_{15}&=-\Col_3' \, ,\\
	\Col_2\mathscr{F}_{15}&=-\frac{\Nc}{2}\Col_2' \, ,\\
	\Col_3\mathscr{F}_{15}&=- \frac{1}{2} \Col_1'-\frac{\Nc}{4}\Col_3'-\frac{\Nc}{4}\Col_4' \, ,\\
	\Col_4\mathscr{F}_{15}&= \left(\frac{1}{\Nc}-\frac{\Nc}{4}\right)\Col_3' - \frac{\Nc}{4}\Col_4' \, .
\end{align}

We thus see that the linear transformation which describes the modification of
the Born-color structure caused by the soft-gluon exchange between two partons
(in this case incoming quark and outgoing gluon) is given by
\begin{equation}
	\mathcal{F}_{15}=
\begin{pmatrix}
	0 & 0 & -\frac{1}{2} & 0\\
	0 & -\frac{\Nc}{2} & 0 & 0\\
	-1 & 0 & -\frac{\Nc}{4}& \frac{1}{\Nc}-\frac{\Nc}{4} \\
	0 & 0 & -\frac{\Nc}{4}& -\frac{\Nc}{4} \, ,
\end{pmatrix}
\label{eq:exmpl_15}
\end{equation}
which represents the color part of the soft anomalous dimension matrix.
In other words, the matrices $\mathcal{F}_{ij}$ are defined such that:
\begin{equation}
	\Col\mathscr{F}_{ij} = \Col'\mathcal{F}_{ij},
\end{equation}
where $\Col=(\Col_1,\Col_2,\Col_3,\Col_4)$ and $\Col'=(\Col_1',\Col_2',\Col_3',\Col_4')$.

The result presented in Eq.~(\ref{eq:exmpl_15}) coincides with the one already
presented in Ref.~\cite{Sjodahl:2008fz}, where one finds also the results for
all other $(i,j)$ pairs.
In this way we calculated the $\mathcal{F}_{15}$ matrix for the graph shown,
as well as those for other Wilson web graphs,
as discussed in Sections~\ref{sec:4softcalc} and~\ref{sec:softdim}.


{\footnotesize{
\providecommand{\href}[2]{#2}\begingroup\raggedright\endgroup
}}

\end{document}